\documentclass[prb,twocolumn,showpacs,amsmath,amssymb]{revtex4}

\usepackage{graphicx}

\begin{document}

%\preprint{ver 06.10.2009}

\title{Quantized antiferromagnetic spin waves in the molecular Heisenberg ring CsFe$_8$}

\author{J. Dreiser}
\email{jan.dreiser@physik.uni-freiburg.de}

\author{O. Waldmann}
\email{oliver.waldmann@physik.uni-freiburg.de} \affiliation{Physikalisches Institut, Universit\"at Freiburg, D-79104
Freiburg, Germany}

\author{C. Dobe}
\author{G. Carver}
\author{S. T. Ochsenbein}
\author{A. Sieber}
\author{H. U. G\"udel}
\affiliation{Department of Chemistry and Biochemistry, University of Bern, 3012 Bern, Switzerland}

\author{J. van Duijn}
\author{J. Taylor}
\affiliation{ISIS Facility, Rutherford Appleton Laboratory, Chilton, Didcot, Oxfordshire OX11 0QX, United Kingdom}

\author{A. Podlesnyak}
\altaffiliation{Present address: Spallation Neutron Source, Oak Ridge National Laboratory, Oak Ridge, Tennessee 37831, USA}
\affiliation{Paul Scherrer Institute, CH-5232 Villigen, Switzerland}

\date{\today}% It is always \today, today,

\begin{abstract}
We report on inelastic neutron scattering (INS) measurements on the molecular spin ring CsFe$_8$, in which eight
spin-5/2 Fe(III) ions are coupled by nearest-neighbor antiferromagnetic Heisenberg interaction. We have recorded
INS data on a non-deuterated powder sample up to high energies at the time-of-flight spectrometers FOCUS at PSI and MARI at ISIS,
which clearly show the excitation of spin waves in the ring. Due to the small number of spin sites, the
spin-wave dispersion relation is not continuous but quantized. Furthermore, the system exhibits a gap between
the ground state and the first excited state. We have modeled our data using exact diagonalization of a
Heisenberg-exchange Hamiltonian together with a small single-ion anisotropy term. Due to the molecule's
symmetry, only two parameters $J$ and $D$ are needed to obtain excellent agreement with the data. The results
can be well described within the framework of the rotational-band model as well as antiferromagnetic spin-wave
theories.
\end{abstract}


%# 75.50.Xx
%Molecular magnets
%# 75.10.Jm
%Quantized spin models (magnetism)
%# 78.70.Nx
%Neutron inelastic scattering (condensed matter)

\maketitle

\section{\label{sec:intro} Introduction}

In recent years, large progress has been made in the understanding and control of mesoscopic magnetic systems, which is driven by the interest in fundamental research as well as potential applications in molecular devices.\cite{RochaNM05,BoganNM08,SchlePRL08,BertaN08,ManniNM09} Dimensionality plays a key role here: magnetic confinement in one or more dimensions leads to strong changes in the density of states, i.e., quantization effects occur,\cite{CloizPR62,WeihoPRB93,HendrPRB93,MathiPRL98,JorziPRL02,PodbiPRL06,ToppPRB08} which become more pronounced with lower dimensionality. Theoretical work on spin waves in low-dimensional antiferromagnetic (AFM) systems reaches back to the early 1950s. While low-temperature thermodynamical properties in two and three dimensions are well described by a conventional spin-wave approach, the case of one dimension (1D)
is more difficult due to divergencies in the sublattice magnetization.\cite{AnderPR52,KuboPR52} These divergencies are deeply
rooted in the fact that, by construction, spin-wave theory (SWT) starts out from the assumption of a long-range
ordered ground state, and their appearance provides a hint to the nature of the ground state, i.e., whether it
is disordered (divergency) or long-range ordered (no divergency). However, despite these subtleties it has been
found that also for disordered systems SWTs can produce results of high accuracy, e.g., for the ground-state
energy.\cite{AnderPR52} Furthermore, the divergencies could subsequently be removed by introducing constraints on the
sublattice magnetization in the so-called modified SWTs \cite{OguchPR60, TakahPRL87,TakahPRB89,HirscPRB89,TangPRB89,RezenPRB90} and finite-size SWTs.\cite{ZhongEL93,LavalPRL98,TrumpPRB00} This establishes a significant improvement, but nevertheless the applicability of SWTs to disordered systems remains a topic of current research. Formerly, the main focus of the theoretical and experimental activities has been on systems of infinite
extension,\cite{ItohPRL95,IvanoPRB92} and it has only been in recent years that research on finite, or zero-dimensional
(0D), systems has been intensified. In this context, the topic of quantized spin waves in mesoscopic spin
clusters has been investigated.\cite{HendrPRB93,WaldmPRL03,ChaboPRB04,OchseEL07,OchseCEJ08,WiesePRL08,YamamPRL02,HoriPRB03}

A good physical realization of such mesoscopic spin clusters are molecular nanomagnets.\cite{Dante06} They have been shown to be a
vast experimental playground enabling the study of precisely defined clusters with chemically engineered exchange
couplings and magnetic anisotropy, or topology. So far, a variety of mesoscopic and quantum phenomena such as tunneling
of the magnetization and stepwise change in the magnetization have been demonstrated (for reviews see Refs.
\onlinecite{BarbaJoMaMM99,GChMB00,GatteACIE03,Dante06,Werns07}). A particularly intriguing type of spin clusters are AFM Heisenberg rings, in which
a number of magnetic metal ions are arranged in the shape of a ring with nearest-neighbor AFM Heisenberg couplings. The
Hamiltonian describing their magnetism is given in its simplest form by
\begin{eqnarray}
\label{eqn:spinhJ}
\hat H = -J \left( \sum_{i=1}^{N-1} \hat{\bf{s}}_i \hat{\bf{s}}_{i+1} + \hat{\bf{s}}_N \hat{\bf{s}}_1  \right),
\end{eqnarray}
with $J<0$ the magnetic coupling strength, $N$ the number of magnetic sites, and $\hat{\bf{s}}_i$ the spin
operator of the $i$th ion with spin $s$. Their magnetism and spin dynamics have intensely been studied by a host
of different experimental techniques;\cite{TaftJotACS94,JuliePRL99,CorniPRB99,SlageC-AEJ02,CarrePRB03,PilawPRB05} the recent experimental work on a Cr$_8$ spin-$3/2$
ring has probably most clearly demonstrated the nature of the elementary excitations.\cite{WaldmPRL03}

On the theory side, the rotational-band model (RBM) has been established,\cite{Ander84,BernuPRL92,BernuPRB94,ChiolPRL98,SchnaPRB00,WaldmPRB01,Lhuil05} and there has been striking
experimental evidence that this picture is a good description for a variety of finite, bipartite AFM spin
clusters with varying topologies \cite{GuidiPRB04,CaciuPRB05,WaldmCCR05,SantiPRB05,WaldmIC06,GuidiPRB07} as well as some frustrated
systems,\cite{SchnaEL01,GarlePRB06,WaldmPRB07b} but of course also exceptions exist.

The RBM classifies the lowest-lying magnetic excitations into two bands with a parabolic energy dependence on
$S$, the $L$ and $E$ band ($S$ is the total spin quantum number). Physically, the $L$ band can be interpreted as
rotations of the two AFM sublattices and is, so-to-say, a precursor of a long-range N\'eel-ordered ground state
in the infinite lattice (it is also known as the tower of states or quasi-degenerate joint states \cite{BernuPRL92,BernuPRB94}). In
contrast, the $E$-band states can be associated with quantized spin-wave excitations.\cite{MuellePRB81,WaldmPRB01} For both bands, the
energies are proportional to $S(S+1)$, but the $E$ band consists of several sub-bands which are shifted up in
energy by constant offsets. For $S \geq 1$, this can be summarized as
\begin{eqnarray}
\label{eqn:ESk}
E(S,k) = \frac{1}{2} \epsilon(\pi) S(S+1) + \epsilon(k) - \epsilon(\pi),
\end{eqnarray}
with $k=0$ or $\pi$ for the $L$ band, where $\epsilon(k)$ may be regarded as the finite-size version of a
spin-wave dispersion relation, and $k$ would become a wave-vector in an infinite system.

There are several reasons for classifying the low-lying excitations into $L$ and $E$ band: the
spectrum of a conventional SWT exhibits zero-energy (Goldstone) modes, with wave vector $k=0,\pi$, reflecting the (assumed or factual) long-range order, which in the context of extended magnets are included in the notation of spin waves or
magnons. However, in a finite system these $k=0,\pi$ modes exhibit a gap with respect to the ground-state energy, forming the $L$ band. These excitations are hence of different physical significance than the higher-lying magnon states ($E$ band). The different nature of the $L$ and $E$ band is further manifested by the fact, that only the $k=0,\pi$ modes are responsible for the diverging sublattice magnetizations in the conventional SWTs, that the $L$- and $E$-band states have different sublattice lengths, and that the scaling with $N$ is generally different. In addition, this distinction is supported by a selection rule for the inelastic neutron scattering (INS) intensities, which at low temperatures allows only INS transitions between $L$ band states or
from $L$ band to $E$ band states.\cite{WaldmPRB01} This key characteristic has been experimentally demonstrated in great detail.\cite{WaldmPRL03} Owing to the obviously different physical properties in small systems, the $L$ band states should be distinguished from the spin waves or magnons.

In the above description, the lowest excited state of an AFM Heisenberg ring is the $S=1$ state of the $L$ band, which
to a good approximation has a gap of $\Delta_c = 4 |J| /N$ to the $S=0$ ground state.
Obviously, $\Delta_c$ vanishes for $N \rightarrow \infty$; it is thus a property of finite rings. In this context, Haldane's conjecture
\cite{HaldaPRL83} should also be mentioned. It predicts that integer-spin 1D chains exhibit a gap in the energy spectrum
while half-integer 1D spin chains do not. In small rings, however, there is numerical evidence that the gap is largely
dominated by $\Delta_c$ without discriminating integer or half-integer spin.\cite{SchnaPRB00a,WaldmPRB01,EngelPRB06} This criterion, in fact, may be regarded
as a definition of what we mean by ``small" or ``0D". A further consequence of the small size is obviously that the spin waves
become quantized, as there can be at most as many spin-wave modes as there are spin sites in the cluster (the number of
spin waves is slightly smaller than $N$, for even AFM rings it is $N-2$).

\begin{figure}
\includegraphics[width=5.0cm]{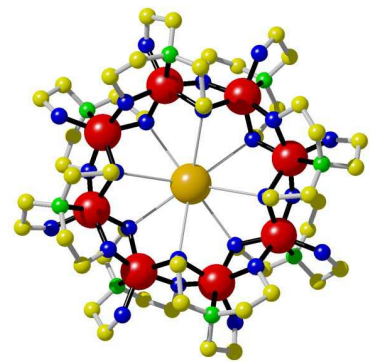}% Here is how to import EPS art
\caption{\label{fig:mol} (Color online) Crystal structure of CsFe$_8$ [with Fe as
dark gray spheres (red online) and H atoms omitted].}
\end{figure}

In this work we investigate the molecule [CsFe$_8$\{N(CH$_2$CH$_2$O)$_3$\}$_8$]Cl, or CsFe$_8$ in short
(Fig.~\ref{fig:mol}). By a variety of experimental techniques, such as high-field torque magnetometry and INS,
the CsFe$_8$ molecule has been demonstrated to be a good model system for an AFM Heisenberg
ring.\cite{SaalfACIEiE97,WaldmIC01} It is characterized by eight Fe(III) ions ($s = 5/2$), which form an almost
planar ring with a crystallographic $C_4$ symmetry axis ($z$ axis) perpendicular to the wheel plane. The
CsFe$_8$ molecule was shown to be very well described by a generic spin Hamiltonian with dominant
nearest-neighbor AFM Heisenberg couplings plus an uniaxial single-ion anisotropy term,
\begin{eqnarray}
\label{eqn:spinh}
\hat H = -J \left( \sum_{i=1}^{N-1} \hat{\bf{s}}_i \hat{\bf{s}}_{i+1} + \hat{\bf{s}}_N \hat{\bf{s}}_1  \right) + D \sum_{i=1}^N \hat{s}_{i,z}^2.
\end{eqnarray}
All the previous experimental data can consistently be described, with an accuracy margin of few percents, with
the values $J = -1.79$~meV and $D = -0.05$~meV. However, all these experiments were only sensitive to the $L$ band states. Here we extend the studies to higher energies, allowing for an investigation of the $E$-band states. The presented INS data will experimentally confirm the expected quantized spin-wave modes, complete the picture of the excitation modes in the CsFe$_8$ molecule, and thus establish the only second example of a full experimental test of the RBM in a small spin cluster. Furthermore, we will describe the $E$-band excitations by different SWTs, comparing their performance to each other. This provides insight into their merits and
drawbacks when applied to mesoscopic spin clusters.

The paper is organized as follows: in the experimental Sec. \ref{sec:exp} the INS spectra and $S(Q,\omega)$ plots are presented. Further, in Sec. \ref{sec:analysis} the INS experiments are described in terms of the microscopic spin Hamiltonian Eq.~\eqref{eqn:spinh}. In Sec. \ref{sec:discuss} the $E$-band excitations are discussed in a more general fashion, revealing their spin-wave character and testing different SWTs. Finally, Sec. \ref{sec:concl} contains the summary and concludes the paper.

\section{\label{sec:exp}Experiments}

A powder sample of CsFe$_8$ was prepared according to literature procedures using non-deuterated starting
materials.\cite{SaalfACIEiE97} INS data were recorded on the direct time-of-flight spectrometer FOCUS at the
Paul-Scherrer Institute (Villigen, Switzerland) using neutrons with incident wavelengths $\lambda = 2.3$ and 3.8~{\AA},
and the MARI spectrometer at the pulsed neutron source ISIS (Didcot, U.K.) using neutrons with incident energies of $E_i
= 15$ and 25~meV. The experimental resolutions at the elastic line were 1.13 and 0.28~meV for the FOCUS, and 0.39 and
0.72~meV for the MARI experiments. If not stated otherwise, the data were summed over all detector banks, and the
neutron-energy loss side is shown.

\begin{figure}
\includegraphics[width=7.8cm]{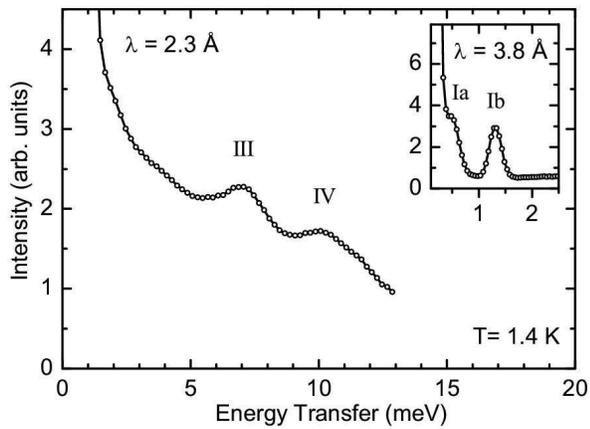}% Here is how to import EPS art
\caption{\label{fig:focus}
INS spectra taken on FOCUS at a temperature of $T=1.4~$K and wavelengths $\lambda = 2.3$~{\AA} (main panel) and
3.8~{\AA} (inset).}
\end{figure}

\begin{figure}
\includegraphics[width=7.8cm]{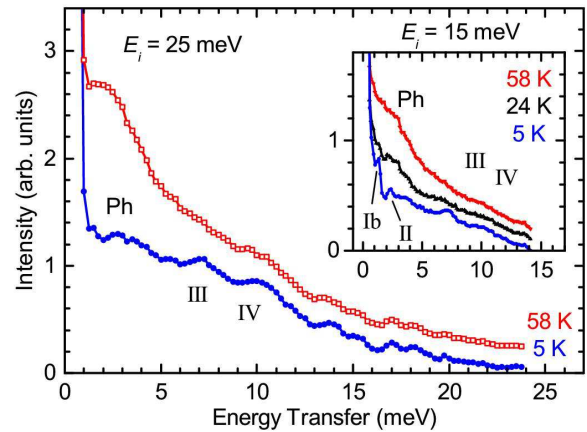}% Here is how to import EPS art
\caption{\label{fig:tdep}(Color online)
Temperature dependence of the INS spectra as recorded on MARI for incident neutron energies of 25~meV (main panel) and
15~meV (inset). Intensity was summed over all detector banks. }
\end{figure}

Figure~\ref{fig:focus} presents the 2.3~{\AA} FOCUS data taken at a temperature of $T = 1.4$~K. Clearly, two cold peaks
at energy transfers of approximately 7~meV (peak III) and 10~meV (peak IV) are visible. The 3.8~{\AA} spectrum, shown in the
inset of Fig.~1, reveals two additional low-energy peaks at energy transfers of about 0.5~meV (peak Ia) and 1.3~meV
(peak Ib). Figure~\ref{fig:tdep} shows INS spectra recorded on MARI at temperatures of 5-58~K. The peaks
Ib, III, and IV can again be observed, but thanks to the larger energy range and higher resolution, additional features
become visible. In particular, a further peak at about 2.3~meV (peak II) is resolved in the 15~meV data (inset of
Fig.~\ref{fig:tdep}). An inspection of the temperature dependence of these four peaks suggests that they are cold
transitions and of magnetic origin. Additional features can be observed, such as the broad feature at about 2.5~meV (feature marked by ``Ph") in the
high-temperature data and the many features at energy transfers above $13$~meV. Their temperature dependence suggests a
phononic origin.

The observed low-energy peaks Ia, Ib, and II are fully consistent with the previous investigations on this molecule, in
particular, the INS experiments.\cite{WaldmPRB06,WaldmPRL05} Peaks Ia and Ib correspond to the transitions from the
$S=0$ ground state to the first-excited $S=1$ level, which is split by magnetic anisotropy into its components $M=0$
and $M=\pm1$ ($M$ is the magnetic quantum number of the total spin $S$). Peak II corresponds to the transition from the lowest excited state to the first excited $S=2$ level. The transitions III and IV were hitherto not observed; their observation and
identification as spin waves form the main body of this work. In the following, we will analyze the data more carefully
in order to demonstrate the magnetic nature of the peaks I to IV, and the lattice origin of the other features in the
spectra. In doing so, we will focus on the two peaks III and IV (as the magnetic peaks at lower energies are already well understood).

\begin{figure*}
\includegraphics[width=14cm]{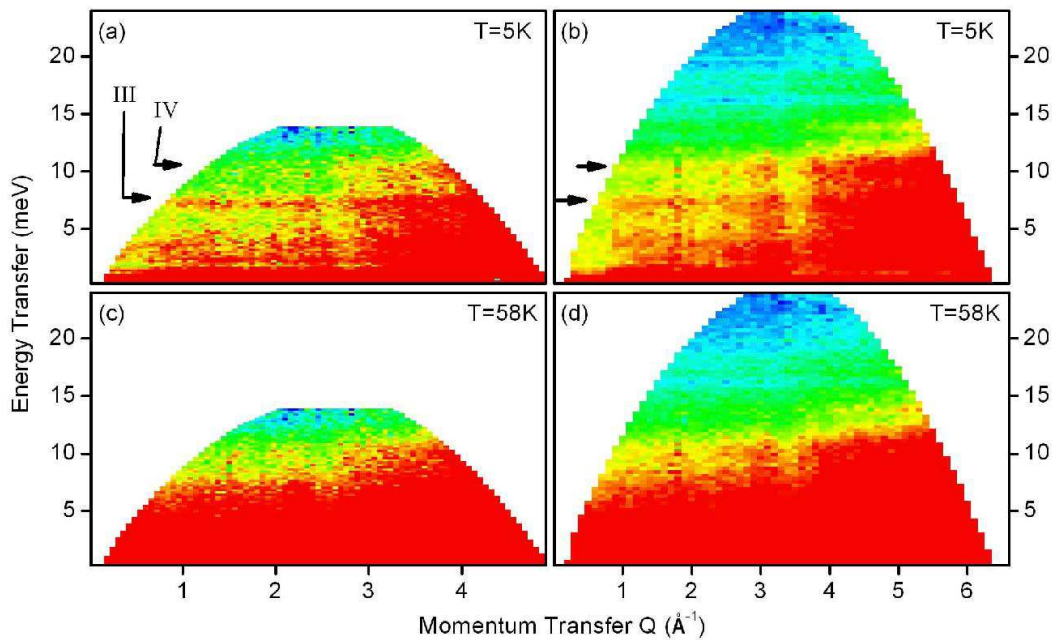}% Here is how to import EPS art
\caption{\label{fig:sqw} (Color online) $S(Q,w)$ plots for the temperatures $T= 5$~K (upper plots) and 58~K (lower
plots) for the two incident neutron energies $E_i=15$~meV (left) and $25$~meV (right) as measured on MARI. The intensity is
color-coded (blue = low, red = high intensity); the color scale is identical for each of the four plots.}
\end{figure*}

First, we analyze the full $S(Q,\omega)$ dependence as presented in Fig.~\ref{fig:sqw} for the
temperatures 5 and 58~K and settings $E_i=15$ and $25$~meV. The $Q$ dependence of the scattering
intensity allows a clear differentiation: the magnetic scattering is determined by the magnetic form factors of the
involved metal ions and interference factors reflecting their spatial arrangement,\cite{WaldmPRB03} it is hence most
significant at lower $Q$ values. In contrast, the phonon scattering intensity increases with $Q$ as $Q^2$. In the
low-temperature data, peaks III and IV (marked by black arrows) are visible in the whole $Q$ range, in particular, at
small $Q$ values, which strongly supports a magnetic origin of these peaks. In the high $Q$ regime the phonon-induced
scattering clearly dominates. The phonon scattering also increases strongly with temperature, as demonstrated by the 58~K data, where scattering is dominated by phonons at all momentum transfers.

\begin{figure}
\includegraphics[width=8cm]{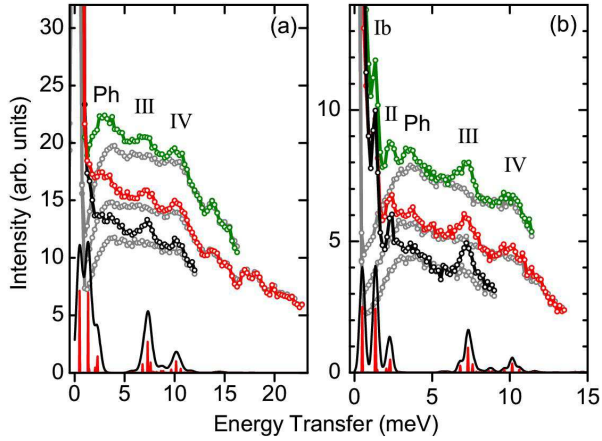}% Here is how to import EPS art
\caption{\label{fig:qslice}
(Color online) $Q$ slices for (a) the $E_i=25$~meV and (b) the $E_i=15$~meV data at $T = 5$~K. In panel (a), the lowest (black), middle
(red), and upper (green) curves correspond to $Q \in [1.0,2.5[$, [2.5,4.0[, and
[4.0,5.5]~{\AA}$^{-1}$, and in panel (b) to $Q \in [1.0,2.0[$, [2.0,3.0[, and
[3.0,4.0]~{\AA}$^{-1}$. The curves have been offset in order to improve visibility.  The gray curves represent the
Bose-factor corrected 58~K data as discussed in the text. The solid lines at the bottom are calculated INS spectra obtained
with the best-fit parameters. The spikes (red) mark the exact position and relative scattering
strengths of the individual INS transitions, the solid (black) curve was obtained by convolution with a Gaussian accounting for the experimental resolution.}
\end{figure}

Figure~\ref{fig:qslice} shows cuts along the energy-axis of the 5~K spectra for three different regimes of momentum
transfer (black, red, and green circles). It reveals that the intensity of the peaks Ib, II, III, and IV roughly stays
constant with $Q$. It might be surprising that these peaks exhibit significant intensity even at $Q$ values as large
as approximately 4~{\AA}, where the magnetic form factors have already dropped considerably; we attribute this to a strong
incoherent scattering in our non-deuterated sample smearing out the $Q$ dependence. These plots further demonstrate that
the feature at approximately 2.5~meV (compare with Fig.~\ref{fig:focus}) must indeed be a phonon.

\begin{figure}
\includegraphics[width=8cm]{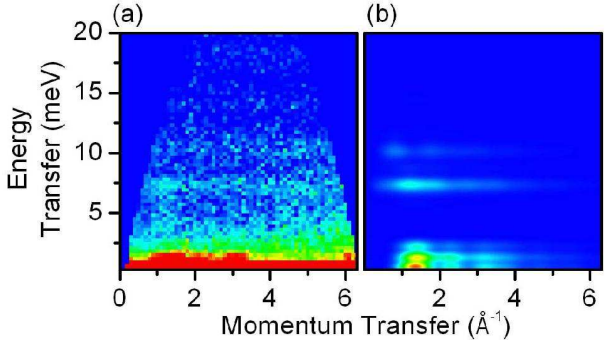}% Here is how to import EPS art
\caption{\label{fig:simsqw}
(Color online) (a) $S(Q,\omega)$ plot of the $E_i=25$~meV MARI data at $T = 5$~K after subtraction of the phonon background as
described in the text. (b) $S(Q,\omega)$ plot ($T = 5$~K) as calculated from the spin Hamiltonian
\eqref{eqn:spinh} with the parameters $J=-1.79$~meV and $D=-0.05$~meV (a Gaussian linewidth of 0.72~meV was used).}
\end{figure}

In Fig.~\ref{fig:qslice}, the gray curves were obtained by scaling the $T = 58$~K data by the Bose factor
$[1-\exp(-\hbar \omega / k_B T)]^{-1} $, which governs the temperature dependence of phonon scattering.\cite{Shira02}
Considering that at high temperatures the magnetic scattering intensity is strongly reduced as it spreads out over
essentially all energies, this curve should provide an estimate of the phonon background. A comparison with the 5~K data nicely
confirms the magnetic nature of peaks III and IV, and also proves that the scattering at higher energies above 12~meV
is purely phononic. The data after correction by the Bose factor is shown in Fig.~\ref{fig:simsqw}(a) as a
$S(Q,\omega)$ plot for $E_i=25$~meV, $T=5$~K. The two peaks III and IV are clearly visible, proving their magnetic origin.

The linewidths of the low-energy peaks Ia, Ib, and II lie within experimental accuracy determined by the
instrumental resolution. In contrast, the linewidths of peaks III and IV are considerably broadened, as is most
clearly seen in Fig.~\ref{fig:qslice}(b). However, our analysis will reveal that they consist of several
close-lying transitions. Finally, it is interesting to note that in the energy regime above peak IV, i.e., above
approximately 13~meV, no significant magnetic scattering intensity could be observed. In this regime, the scattering is
entirely of phononic origin, as is evident from Fig.~\ref{fig:simsqw}(a) and even more so from
Fig.~\ref{fig:qslice}(a).

\section{\label{sec:analysis}Analysis}

\begin{figure}
\includegraphics[width=8cm]{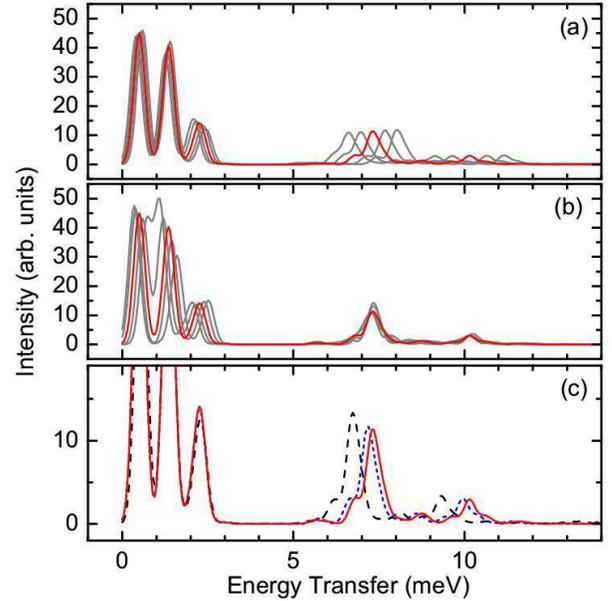}% Here is how to import EPS art
\caption{\label{fig:jdep}
(Color online) (a) Dependence of the INS spectrum on $J$, which is varied by a maximum of $\pm 10\%$ around a center value of
$J = -1.79$~meV (red curve). The anisotropy is $D = -0.05$~meV. (b) Dependence of the INS spectrum on $D$, which
is varied by a maximum $\pm 50 \%$ around a center value of $D = -0.05$~meV (red curve). The coupling is $J=-1.79$~meV. (c)
Dependence of the INS spectrum on a modulation of the $J$ values along the ring. For all curves the anisotropy
is $D = -0.05$~meV. Red (solid) curve: no modulation, $J_1=J_2 = -1.79$~meV. Blue (short dashed) curve: $J_1 =
-2.14$~meV and $J_2 = -1.45$~meV. Black (long dashed) curve: $J_1 = -2.48$~meV and $J_2 = -1.10$~meV.}
\end{figure}

We have simulated the INS spectrum of Eq.~\eqref{eqn:spinh} by calculating the energies and
wave functions via a sparse-matrix exact numerical diagonalization method and using the formulas of
Refs.~\onlinecite{WaldmPRB03} and \onlinecite{WaldmPRB05a}. In order to find the parameters which yield the best fit to the experimental spectra we have systematically scanned the $J$ and $D$ parameter space, yielding $J = -1.79(5)$~meV and $D = -0.05(7)$~meV. These values are
in excellent agreement with the previous findings.\cite{WaldmPRB06,WaldmPRL05} The dependence of the simulated
INS spectrum on $J$ and $D$ is shown in Figs.~\ref{fig:jdep}(a) and \ref{fig:jdep}(b), respectively. The
low-energy part of the spectrum is affected by both $J$ and $D$. In contrast, the high-energy part is predominantly affected by $J$. Indeed, a 10$\%$~variation in $J$ has a much larger effect than a 50$\%$~variation in $D$. This explains the rather large estimated standard deviation of the $D$ value that we have determined. The simulated INS spectra corresponding to the best-fit values are shown at the bottom of Figs.~\ref{fig:qslice}(a) and \ref{fig:qslice}(b) as solid lines, and as a $S(Q,\omega)$ plot in Fig.~\ref{fig:simsqw}(b); the good agreement with the data is obvious.

\begin{figure}
\includegraphics[width=8cm]{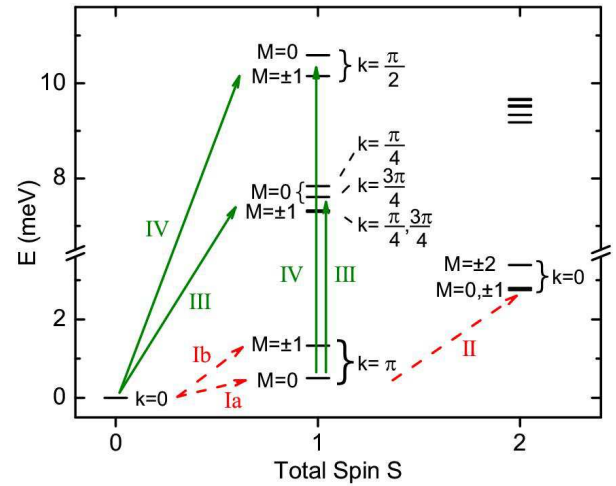}% Here is how to import EPS art
\caption{\label{fig:instrans}
(Color online) Calculated low-energy spectrum of Hamiltonian \eqref{eqn:spinh} with $J = -1.79$~meV and $D =
-0.05$~meV. Only states in the sector $S \leq 2$ are plotted. The energy of the ground state was set to zero.
Due to the anisotropy term, the degeneracy in $M$ of each $S$ multiplet is partially lifted. The relevant INS
transitions are plotted as solid green (dashed red) arrows, marking transitions to the $E$ band ($L$ band). }
\end{figure}

Figure~\ref{fig:instrans} shows the relevant part of the calculated spectrum, detailing the assignment of the observed peaks. The assignment of the lower-energy peaks has already been discussed in Refs.~\onlinecite{WaldmPRB06} and \onlinecite{WaldmPRL05} and in the above. The peaks III and IV correspond to transitions from the $S=0$ ground state to higher-lying $S=1$ levels, which are slightly split due to the
magnetic anisotropy. Accordingly, each of the peaks consists of a number of close-lying transitions, giving rise
to the experimentally observed broadening of the peaks. Furthermore, at the measurement temperature of 5~K, also the
transitions from the first excited state weakly contribute. The INS intensities of the individual transitions
are indicated by the vertical thin lines in Fig.~\ref{fig:qslice}. The calculated contributions to all observed
INS peaks, i.e., the quantum numbers of the involved states, the transition energies, and the scattering
intensities, are detailed in Table~\ref{tab:instrans}.
\begin{table}
\caption{\label{tab:instrans} Calculated contributions to the observed INS transitions. The quantum states are
labeled as $|S,M,k\rangle$, as in  Fig.~\ref{fig:instrans}. The ground state is $|\textrm {gs} \rangle =
|0,0,0\rangle$. $|i \rangle$ and $|f \rangle$ denotes the initial and final state in the transition. $E_f -E_i$
is the excitation energy in units of meV, and $I_{if}$ indicates the transition strength, normalized to the
strongest transition Ia. Transitions with strength $I_{if} < 0.05$ have been omitted.} 
\begin{ruledtabular}
\begin{tabular}{ccccc}
Peak & $|i \rangle$ &  $|f \rangle$ & $E_f - E_i$ & $I_{if}$ \\
\hline
Ia & $|\textrm {gs} \rangle $ & $ |1,0, \pi \rangle$ & 0.50 & 1.00 \\
\hline
Ib & $|\textrm {gs} \rangle $ & $ |1,\pm 1, \pi \rangle$ & 1.34 & 0.97 \\
\hline
II  & $|1,\pm 1, \pi \rangle $ & $ |2,\pm 2, 0 \rangle$ & 2.05 &	0.06  \\
		& $|1,0, \pi \rangle $ & $ |2,0, 0 \rangle$ & 2.25 & 0.12  \\
   	& $|1,0, \pi \rangle $ & $ |2, \pm 1, 0 \rangle$ & 2.30 & 0.20  \\
\hline
III & $|1,0, \pi \rangle $ & $ |1,0,  3 \pi/4 \rangle$ & 6.83 & 0.09 \\ 
		& $|\textrm {gs} \rangle $ & $ |1,\pm 1, 3 \pi/4 \rangle$ & 7.30 & 0.33  \\
    & $|\textrm {gs}  \rangle $ & $ |1,\pm 1,  \pi /4 \rangle$ & 7.33 & 0.05  \\
    & $|\textrm {gs}  \rangle $ & $ |1,0, 3 \pi/4 \rangle$ & 7.61 & 0.13  \\
\hline
IV 	& $|\textrm {gs}  \rangle $ & $ |1, 0, \pi/2 \rangle$ & 10.59 & 0.05  \\
   	& $|\textrm {gs}  \rangle $ & $ |1,\pm 1, \pi/2 \rangle$ & 10.15 & 0.14  \\
\end{tabular}
\end{ruledtabular}
\end{table}
The model Hamiltonian \eqref{eqn:spinh} assumes a perfect symmetry of the ring (i.e., an eight-fold symmetry
axis). However, though it is close to S$_8$, the molecular symmetry of CsFe$_8$ is $C_4$. Hence, weak deviations exist raising the question of how important they are for the magnetism in CsFe$_8$. This point is relevant for the understanding and explanation of phenomena such as the field-induced spin-Jahn-Teller effect \cite{WaldmPRL06,SonciPRL07,LantePRB09} or the significance of possible Dzyaloshinski-Moriya (DM)
interactions.\cite{CintiTEPJB02} In this context, the present experiment is not sensitive to such small effects as a DM interaction or a
biaxial anisotropy term ($E$ term), but a variation in the exchange-coupling constants along the ring, e.g., in
a $J_1$-$J_2$ alternating fashion, can be tested. Indeed, from general arguments it is expected that the
$L$ band states are rather insensitive to such a variation (as their energies are affected only in second order,
see Ref.~\onlinecite{WaldmIC01}). However, this is not the case for the $E$ band or spin-wave states, respectively,
which hence provide a better means for a check. A simulation of the INS spectrum at 5~K with a modulation of the exchange-coupling constant
of up to $\pm 40 \%$ around its best-fit value is shown in Fig.~\ref{fig:jdep}(c). As expected, the low-energy
part is not much affected but the spin-wave excitations at 7 and 10~meV have significantly moved toward lower energies. By comparison with the experimental widths of the peaks we conclude that a modulation of the exchange-coupling constants is smaller than $\pm 20\%$
in CsFe$_8$.

\section{\label{sec:discuss}Discussion}

In the previous sections the following points have been made: peaks III and IV are of magnetic origin, they are well
reproduced by the generic spin Hamiltonian \eqref{eqn:spinh}, they correspond to transitions from the ground
state to excited $S=1$ states, and there is negligible magnetic scattering at energies above peak IV. In
this section we will discuss some implications of these findings.

As presented in Sec.~\ref{sec:intro}, the low-energy spectrum of AFM Heisenberg spin rings can be described by the
RBM, which subdivides the spectrum into the $L$ and $E$ band; all other states are denoted collectively as the
quasi-continuum. The $L$ band is formed by the lowest spin multiplet in each sector of total spin $S$. For CsFe$_8$, the
transitions within the states of the $L$ band were already investigated intensely before;\cite{WaldmPRL05,WaldmPRB06} in the present
work they are observed as peaks Ia, Ib, and II. The $E$ band consists of the $N-2$ next higher-lying spin
multiplets in each spin sector $S\geq1$. These states are exactly those which are observed as the two peaks III and
IV in our INS data. Only two broad INS peaks appear because of the (near) degeneracy of some states,
which are due to the ring symmetry and the two-sublattice structure in the AFM ring, as will also become clear
in the following. The RBM also predicts a selection rule for the matrix elements, which in fact is a key aspect
of the RBM as it ultimately justifies the distinction of the states into the $L$ band, $E$ band, and
quasi-continuum. It states that, starting from the ground state, only transitions to states of the $L$ or $E$
band have significant INS intensities, while transitions into the quasi-continuum are negligibly weak. Thus, at
low temperatures no magnetic INS intensity should be observed at energies above the $E$ band or peak IV,
respectively, exactly as observed in our experiment. Therefore, in conjunction with the previous
work,\cite{WaldmPRL05,WaldmPRB06} which confirmed the $L$ band in great detail, the data presented here provides a thorough
confirmation of the RBM for CsFe$_8$. This is only the second example where this has been achieved (the first
being on the Cr$_8$ molecule, Ref.~\onlinecite{WaldmPRL03}).

The connection between $E$ band and quantized AFM spin waves has been discussed before in the literature
to some extent.\cite{MuellePRB81,BernuPRL92,BernuPRB94,WaldmPRB01,Lhuil05} In the following it will be made explicit by applying various SWTs available in the
literature to a small AFM Heisenberg ring. The analysis will also give insight into the strengths and
weaknesses of these methods when applied to small spin clusters. As it was shown in Sec.~III, the energies of
peaks III and IV do not depend much on the magnetic anisotropy. Hence, it is ignored in the calculations in this section, i.e., Eq.~\eqref{eqn:spinhJ} is used. Excitation energies are given relative to the ground-state energy $E_g$.

It is useful to first discuss the symmetries of the model and Fig.~\ref{fig:instrans}. The AFM Heisenberg ring
exhibits spin-rotational symmetry, which gives rise to quantum numbers $S$ and $M$, and a $D_N$ spatial
symmetry. In a pure Heisenberg model the states are degenerate with respect to $M$. In CsFe$_8$ this degeneracy
is lifted by the magnetic anisotropy, as indicated in Fig.~\ref{fig:instrans}, but the splitting is weak for the
higher-lying states, and a classification in terms of spin multiplets or $S$, respectively, is useful. As regards
the spatial symmetry, it is convenient to relax it to $C_N$. The quantum numbers, which we will call shift
quantum number $q$, are related to a shift operator $\hat T | q \rangle = \exp(\textrm i q 2 \pi /N) |q \rangle$
with $q= 0, 1, ..., N-1$. Alternatively, one may use ``wave vectors" $k = 2 \pi q /N + \text{const}$, where we chose
const such that for the ground state $k=0$. The actual symmetry, however, is $D_N$, which requires that states
with $q$ and $N-q$, or $k$ and $-k$, respectively, are degenerate. We now turn to Fig.~\ref{fig:instrans}. The
$S=0$ sector embraces the ground state with $k=0$. The remaining $S=0$ levels are high up in energy and are part
of the quasi-continuum (not shown in Fig.~\ref{fig:instrans}). In the $S=1$ sector, the (three) lowest states
correspond to the first-excited spin multiplet of the $L$ band, for which $k=\pi$. This multiplet appears split
in Fig.~\ref{fig:instrans} because of the magnetic anisotropy. The next $N-2 = 6$ higher-lying spin multiplets
with $k=\pm \frac{\pi}{4},\pm \frac{\pi}{2},\pm \frac{3\pi}{4}$ form the $E$ band. Their counting is difficult
in Fig.~\ref{fig:instrans} because of the anisotropy splitting and the degeneracies. This pattern of states
continues for $S>1$ (not shown in Fig.~\ref{fig:instrans}).

\begin{figure}
\includegraphics[width=7.8cm]{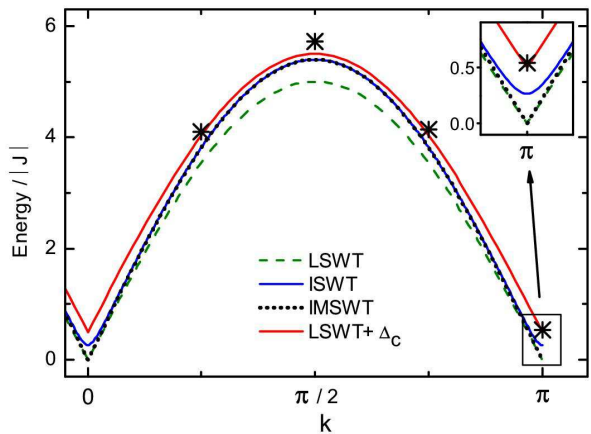}% Here is how to import EPS art
\caption{\label{fig:swt}
(Color online) Spin-wave excitation energies as calculated by exact numerical diagonalization (stars) normalized to $|J|$ and the
different SWTs (solid, dashed and dotted lines) as discussed in the text. The abbreviations
are LSWT: linear SWT, ISWT: interacting SWT, IMSWT: full-diagonalization
interacting modified SWT, LSWT + $\Delta_c$: linear SWT plus classical gap $\Delta_c$. }
\end{figure}

Alternatively to the energy-vs-$S$ representation, the structure of the energy spectrum may be discussed in an
energy-vs-$k$ representation, which for an infinite system would be the dispersion relation. However, in a small
system, with no periodicity along an extended axis, $k$ in general does not correspond to a ``real" wave vector
(although it can approach one for $N \rightarrow \infty$ in the case of a ring). Furthermore, $k$ may
assume only a set of finite values, and in consequence, the allowed energies are discrete values,
yielding quantized spin waves in our context. All available SWTs can easily be applied to small spin clusters if
one chooses to work in real space and not momentum space (see Refs.~\onlinecite{WaldmPRB07b} and \onlinecite{C'epPoTPS05}). Generally, SWTs work with
bosonic excitations from the ground state, the magnons or spin waves. The one-magnon excitations correspond to
the $L$- and $E$-band states in the $S=1$ sector. Other states may, in principle, be generated by multiple-magnon
excitations, but because of the technical difficulties SWT is rarely developed to this stage. In other words,
the SWTs provide us with the energies of the $L$- and $E$-band states in the $S=1$ sector or the values
$\epsilon(k)$ in Eq.~\eqref{eqn:ESk}, respectively. The dependence on $S$ is essentially out of their reach. The
excitation energies of these states are plotted as a function of $k$ in Fig.~\ref{fig:swt} for different SWTs as well as the exact numerical result.

In the early 1950s, the (conventional) linear and interacting SWTs were established.\cite{AnderPR52,OguchPR60}
Low-dimensional systems present a challenge, since divergencies appear in 1D (and 0D) due to the fact that in
these SWTs the modes with $k=0$ and $\pi$ are degenerate and have zero energy (Goldstone modes). The $L$ band is related to these
modes; hence the gap between the $S=0$ ground state and the first-excited $S=1$ state (singlet-triplet gap),
which in small AFM systems is always present, is zero, i.e., can never be reproduced by these SWTs. Later on,
the modified SWTs\cite{TakahPRB89,HirscPRB89,TangPRB89} and Schwinger-boson mean-field theory
(SBMFT)\cite{AuerbPRL88,SarkePRB89} were introduced, which eliminate the drawback of a zero singlet-triplet gap,
suggesting their applicability to small spin clusters. Fortunately, for all these theories, linear SWT (LSWT),
interacting SWT (ISWT), linear modified SWT (LMSWT), interacting modified SWT (IMSWT), and SBMFT, analytical
results are available for 1D, which can be directly applied to the AFM ring (various IMSWTs exist,
we have used the full-diagonalization IMSWT of Ref.~\onlinecite{YamamJotPSoJ03}). Interestingly, the SBMFT yields exactly the
same excitation energies as the LMSWT, and also the ground-state energy agree after introduction of a correction
factor.\cite{ArovaPRB88,YamamJotPSoJ03} Hence, there is no need to consider the SBMFT separately in the following. The singlet-triplet gap may be
approximated by $\Delta_c = 4 |J| /N$, which becomes exact in the classical limit.\cite{AnderPR52,WaldmPRB01,Lhuil05} Adding this
gap to the excitation energies of the conventional SWTs is tempting (and may be justified, see
Refs.~\onlinecite{ZhongEL93} and \onlinecite{TrumpPRB00}), but does not improve the ISWT; the result for LSWT, denoted as LSWT+$\Delta_c$, is
discussed below.

The excitation energies as function of $k$ are presented in Fig.~\ref{fig:swt}. The interpretation of the INS
transitions in terms of spin waves is obvious: peak IV at 10~meV corresponds to the one-magnon excitations
with $k = \pm \frac{\pi}{2}$, and peak III at 7~meV is related to excitations of magnons with $k = \pm
\frac{\pi}{4}$ and $\pm \frac{3\pi}{4}$. The ``quantization" of the spin-wave energies due to the system's finite
size can also be regarded as being due to a ``confinement" of the magnons in three-dimensional space. The
dispersion relation as calculated by SWT is also symmetric with respect to $k= \pm \frac{\pi}{4}$ due to the AFM
sublattice structure.

Table~II compiles the ground-state energy $E_g$, singlet-triplet gap $\Delta$, \footnote{We define the singlet-triplet gap in the case of SWTs as $\Delta = \epsilon(\pi)$.} width $\epsilon(\pi/2)-\epsilon(\pi)$ of the excitation spectrum, and mean deviation $\chi^2 = \sum_k [\epsilon(k) - \epsilon_\textrm{exact}(k)]^2 /
(\frac{1}{2} N |J| s)$ from the exact energies as calculated by exact numerical diagonalization and the various
SWTs for the case $N=8, s=5/2$. Inspection of Table~II and Fig.~\ref{fig:swt} shows:

(1) As discussed beforehand the conventional SWTs (LSWT and ISWT) do not produce a singlet-triplet gap while
the modified SWTs do. Yet, they grossly underestimate the gap, by a factor of 2. In contrast, the
classical estimate $\Delta_c$ is off by only 7\%. Hence, although the modified SWTs reproduce the singlet-triplet gap, they do not provide a good description of the low-energy or low-temperature behavior, respectively. Accordingly, LSWT+$\Delta_c$ performs best.

(2) The width $\epsilon(\pi/2)-\epsilon(\pi)$ of the spin-wave excitation spectrum is obtained reasonably well in all considered SWTs, with the
expected trend that the interacting versions do better. Except for the ISWT, the SWTs
underestimate the width. It is noted that the modified SWTs generally predict a smaller width than the
corresponding conventional SWTs. Further, among the SWTs, IMSWT yields the best assessment of the excitations. Notably, the empirical LSWT+$\Delta_c$ calculation performs even better than the IMSWT.

(3) Regarding the ground-state energy, the IMSWT yields the best results, with a remarkably small deviation of only 0.002\%. This implies that also the ground-state wave function is accurately obtained by this method. It would be interesting to verify this in the future.

\begin{table}
\caption{\label{tab:swt} Exact and SWT results for the singlet-triplet gap, energy width, and mean deviation for
a $N=8$, $s=5/2$ ring. }
\begin{ruledtabular}
\begin{tabular}{ccccc}
& $E_g/|J|$ & $\Delta/|J| $ &  $[\epsilon(k=\pi/2)-\Delta]/|J|$ & $\chi^2$ \\
\hline
Exact & -58.1105 & 0.537 & 5.180 & 0  \\
LSWT & -57.2600 & 0 & 5 & 0.133 \\
LMSWT & -57.8049 &  0.248 & 4.758 & 0.115  \\
ISWT & -58.2433 & 0 & 5.396 & 0.093 \\
IMSWT & -58.1095 & 0.268 & 5.129 & 0.065  \\
LSWT+$\Delta_c$&  & 0.5 & 5  & 0.025 \\
\end{tabular}
\end{ruledtabular}
\end{table}
\section{\label{sec:concl}Conclusion}

In conclusion, we have observed by INS the excitation of quantized spin-wave modes in the molecule CsFe$_8$,
i.e., an eight-membered AFM Heisenberg ring. Only a small number of discrete points on the $\epsilon(k)$
spin-wave dispersion curve are allowed due to the confinement of the magnetic excitations to the ring. Further, the
INS data revealed that at low temperatures the magnetic scattering intensity at energies above the most
energetic spin-wave excitation is negligibly small. These findings, combined with the demonstration of the $L$
band in previous works, establish a thorough confirmation of the RBM for CsFe$_8$. We have compared several SWTs
with numerically exact results. The full-diagonalization interacting modified SWT delivers excellent
approximations to the ground-state energy and spin-wave excitations, but the singlet-triplet gap and hence the
low-temperature properties of the ring are only qualitatively reproduced. Regarding the excitation energies the empirical LSWT+$\Delta_c$ calculation performs best.

\begin{acknowledgments}
The authors thankfully acknowledge funding by the Deutsche Forschungsgemeinschaft.
\end{acknowledgments}


\begin{thebibliography}{77}

\expandafter\ifx\csname natexlab\endcsname\relax\def\natexlab#1{#1}\fi
\expandafter\ifx\csname bibnamefont\endcsname\relax
  \def\bibnamefont#1{#1}\fi
\expandafter\ifx\csname bibfnamefont\endcsname\relax
  \def\bibfnamefont#1{#1}\fi
\expandafter\ifx\csname citenamefont\endcsname\relax
  \def\citenamefont#1{#1}\fi
\expandafter\ifx\csname url\endcsname\relax
  \def\url#1{\texttt{#1}}\fi
\expandafter\ifx\csname urlprefix\endcsname\relax\def\urlprefix{URL }\fi
\providecommand{\bibinfo}[2]{#2}
\providecommand{\eprint}[2][]{\url{#2}}

\bibitem[{\citenamefont{Rocha et~al.}(2005)\citenamefont{Rocha, Garcia-suarez,
  Bailey, Lambert, Ferrer, and Sanvito}}]{RochaNM05}
\bibinfo{author}{\bibfnamefont{A.~R.} \bibnamefont{Rocha}},
  \bibinfo{author}{\bibfnamefont{V.~M.} \bibnamefont{Garcia-suarez}},
  \bibinfo{author}{\bibfnamefont{S.~W.} \bibnamefont{Bailey}},
  \bibinfo{author}{\bibfnamefont{C.~J.} \bibnamefont{Lambert}},
  \bibinfo{author}{\bibfnamefont{J.}~\bibnamefont{Ferrer}}, \bibnamefont{and}
  \bibinfo{author}{\bibfnamefont{S.}~\bibnamefont{Sanvito}},
  \bibinfo{journal}{Nature Mater.} \textbf{\bibinfo{volume}{4}},
  \bibinfo{pages}{335} (\bibinfo{year}{2005}).

\bibitem[{\citenamefont{Bogani and Wernsdorfer}(2008)}]{BoganNM08}
\bibinfo{author}{\bibfnamefont{L.}~\bibnamefont{Bogani}} \bibnamefont{and}
  \bibinfo{author}{\bibfnamefont{W.}~\bibnamefont{Wernsdorfer}},
  \bibinfo{journal}{Nature Mater.} \textbf{\bibinfo{volume}{7}},
  \bibinfo{pages}{179} (\bibinfo{year}{2008}).

\bibitem[{\citenamefont{Schlegel et~al.}(2008)\citenamefont{Schlegel, van
  Slageren, Manoli, Brechin, and Dressel}}]{SchlePRL08}
\bibinfo{author}{\bibfnamefont{C.}~\bibnamefont{Schlegel}},
  \bibinfo{author}{\bibfnamefont{J.}~\bibnamefont{van Slageren}},
  \bibinfo{author}{\bibfnamefont{M.}~\bibnamefont{Manoli}},
  \bibinfo{author}{\bibfnamefont{E.~K.} \bibnamefont{Brechin}},
  \bibnamefont{and} \bibinfo{author}{\bibfnamefont{M.}~\bibnamefont{Dressel}},
  \bibinfo{journal}{Phys. Rev. Lett.} \textbf{\bibinfo{volume}{101}},
  \bibinfo{pages}{147203} (\bibinfo{year}{2008}).

\bibitem[{\citenamefont{Bertaina et~al.}(2008)\citenamefont{Bertaina,
  Gambarelli, Mitra, Tsukerblat, M\"uller, and Barbara}}]{BertaN08}
\bibinfo{author}{\bibfnamefont{S.}~\bibnamefont{Bertaina}},
  \bibinfo{author}{\bibfnamefont{S.}~\bibnamefont{Gambarelli}},
  \bibinfo{author}{\bibfnamefont{T.}~\bibnamefont{Mitra}},
  \bibinfo{author}{\bibfnamefont{B.}~\bibnamefont{Tsukerblat}},
  \bibinfo{author}{\bibfnamefont{A.}~\bibnamefont{M\"uller}}, \bibnamefont{and}
  \bibinfo{author}{\bibfnamefont{B.}~\bibnamefont{Barbara}},
  \bibinfo{journal}{Nature} \textbf{\bibinfo{volume}{453}},
  \bibinfo{pages}{203} (\bibinfo{year}{2008}).

\bibitem[{\citenamefont{Mannini et~al.}(2009)\citenamefont{Mannini, Pineider,
  Sainctavit, Danieli, Otero, Sciancalepore, Talarico, Arrio, Cornia, Gatteschi, and Sessoli}}]{ManniNM09}
\bibinfo{author}{\bibfnamefont{M.}~\bibnamefont{Mannini}},
  \bibinfo{author}{\bibfnamefont{F.}~\bibnamefont{Pineider}},
  \bibinfo{author}{\bibfnamefont{P.}~\bibnamefont{Sainctavit}},
  \bibinfo{author}{\bibfnamefont{C.}~\bibnamefont{Danieli}},
  \bibinfo{author}{\bibfnamefont{E.}~\bibnamefont{Otero}},
  \bibinfo{author}{\bibfnamefont{C.}~\bibnamefont{Sciancalepore}},
  \bibinfo{author}{\bibfnamefont{A.~M.} \bibnamefont{Talarico}},
  \bibinfo{author}{\bibfnamefont{M.-A.} \bibnamefont{Arrio}},
  \bibinfo{author}{\bibfnamefont{A.}~\bibnamefont{Cornia}},
  \bibinfo{author}{\bibfnamefont{D.}~\bibnamefont{Gatteschi}}, \bibnamefont{and}
  \bibinfo{author}{\bibfnamefont{R.}~\bibnamefont{Sessoli}},
  \bibinfo{journal}{Nature Mater.}
  \textbf{\bibinfo{volume}{8}}, \bibinfo{pages}{194} (\bibinfo{year}{2009}).

\bibitem[{\citenamefont{des Cloizeaux and Pearson}(1962)}]{CloizPR62}
\bibinfo{author}{\bibfnamefont{J.}~\bibnamefont{des Cloizeaux}}
  \bibnamefont{and} \bibinfo{author}{\bibfnamefont{J.~J.}
  \bibnamefont{Pearson}}, \bibinfo{journal}{Phys. Rev.}
  \textbf{\bibinfo{volume}{128}}, \bibinfo{pages}{2131} (\bibinfo{year}{1962}).

\bibitem[{\citenamefont{Weihong and Hamer}(1993)}]{WeihoPRB93}
\bibinfo{author}{\bibfnamefont{Zheng}~\bibnamefont{Weihong}} \bibnamefont{and}
  \bibinfo{author}{\bibfnamefont{C.~J.} \bibnamefont{Hamer}},
  \bibinfo{journal}{Phys. Rev. B} \textbf{\bibinfo{volume}{47}},
  \bibinfo{pages}{7961} (\bibinfo{year}{1993}).

\bibitem[{\citenamefont{Hendriksen et~al.}(1993)\citenamefont{Hendriksen,
  Linderoth, and Lindg\aa{}rd}}]{HendrPRB93}
\bibinfo{author}{\bibfnamefont{P.~V.} \bibnamefont{Hendriksen}},
  \bibinfo{author}{\bibfnamefont{S.}~\bibnamefont{Linderoth}},
  \bibnamefont{and} \bibinfo{author}{\bibfnamefont{P.-A.}
  \bibnamefont{Lindg\aa{}rd}}, \bibinfo{journal}{Phys. Rev. B}
  \textbf{\bibinfo{volume}{48}}, \bibinfo{pages}{7259} (\bibinfo{year}{1993}).

\bibitem[{\citenamefont{Mathieu et~al.}(1998)\citenamefont{Mathieu, Jorzick,
  Frank, Demokritov, Slavin, Hillebrands, Bartenlian, Chappert, Decanini,
  Rousseaux, and Cambril}}]{MathiPRL98}
\bibinfo{author}{\bibfnamefont{C.}~\bibnamefont{Mathieu}},
  \bibinfo{author}{\bibfnamefont{J.}~\bibnamefont{Jorzick}},
  \bibinfo{author}{\bibfnamefont{A.}~\bibnamefont{Frank}},
  \bibinfo{author}{\bibfnamefont{S.~O.} \bibnamefont{Demokritov}},
  \bibinfo{author}{\bibfnamefont{A.~N.} \bibnamefont{Slavin}},
  \bibinfo{author}{\bibfnamefont{B.}~\bibnamefont{Hillebrands}},
  \bibinfo{author}{\bibfnamefont{B.}~\bibnamefont{Bartenlian}},
  \bibinfo{author}{\bibfnamefont{C.}~\bibnamefont{Chappert}},
  \bibinfo{author}{\bibfnamefont{D.}~\bibnamefont{Decanini}},
  \bibinfo{author}{\bibfnamefont{F.}~\bibnamefont{Rousseaux}}, \bibnamefont{and}
  \bibinfo{author}{\bibfnamefont{E.}~\bibnamefont{Cambril}}, 
  \bibinfo{journal}{Phys. Rev. Lett.}
  \textbf{\bibinfo{volume}{81}}, \bibinfo{pages}{3968} (\bibinfo{year}{1998}).

\bibitem[{\citenamefont{Jorzick et~al.}(2002)\citenamefont{Jorzick, Demokritov,
  Hillebrands, Bailleul, Fermon, Guslienko, Slavin, Berkov, and
  Gorn}}]{JorziPRL02}
\bibinfo{author}{\bibfnamefont{J.}~\bibnamefont{Jorzick}},
  \bibinfo{author}{\bibfnamefont{S.~O.} \bibnamefont{Demokritov}},
  \bibinfo{author}{\bibfnamefont{B.}~\bibnamefont{Hillebrands}},
  \bibinfo{author}{\bibfnamefont{M.}~\bibnamefont{Bailleul}},
  \bibinfo{author}{\bibfnamefont{C.}~\bibnamefont{Fermon}},
  \bibinfo{author}{\bibfnamefont{K.~Y.} \bibnamefont{Guslienko}},
  \bibinfo{author}{\bibfnamefont{A.~N.} \bibnamefont{Slavin}},
  \bibinfo{author}{\bibfnamefont{D.~V.} \bibnamefont{Berkov}},
  \bibnamefont{and} \bibinfo{author}{\bibfnamefont{N.~L.} \bibnamefont{Gorn}},
  \bibinfo{journal}{Phys. Rev. Lett.} \textbf{\bibinfo{volume}{88}},
  \bibinfo{pages}{047204} (\bibinfo{year}{2002}).

\bibitem[{\citenamefont{Podbielski et~al.}(2006)\citenamefont{Podbielski,
  Giesen, and Grundler}}]{PodbiPRL06}
\bibinfo{author}{\bibfnamefont{J.}~\bibnamefont{Podbielski}},
  \bibinfo{author}{\bibfnamefont{F.}~\bibnamefont{Giesen}}, \bibnamefont{and}
  \bibinfo{author}{\bibfnamefont{D.}~\bibnamefont{Grundler}},
  \bibinfo{journal}{Phys. Rev. Lett.} \textbf{\bibinfo{volume}{96}},
  \bibinfo{eid}{167207}  (\bibinfo{year}{2006}).

\bibitem[{\citenamefont{Topp et~al.}(2008)\citenamefont{Topp, Podbielski,
  Heitmann, and Grundler}}]{ToppPRB08}
\bibinfo{author}{\bibfnamefont{J.}~\bibnamefont{Topp}},
  \bibinfo{author}{\bibfnamefont{J.}~\bibnamefont{Podbielski}},
  \bibinfo{author}{\bibfnamefont{D.}~\bibnamefont{Heitmann}}, \bibnamefont{and}
  \bibinfo{author}{\bibfnamefont{D.}~\bibnamefont{Grundler}},
  \bibinfo{journal}{Phys. Rev. B} \textbf{\bibinfo{volume}{78}},
  \bibinfo{eid}{024431} (\bibinfo{year}{2008}).

\bibitem[{\citenamefont{Anderson}(1952)}]{AnderPR52}
\bibinfo{author}{\bibfnamefont{P.~W.} \bibnamefont{Anderson}},
  \bibinfo{journal}{Phys. Rev.} \textbf{\bibinfo{volume}{86}},
  \bibinfo{pages}{694} (\bibinfo{year}{1952}).

\bibitem[{\citenamefont{Kubo}(1952)}]{KuboPR52}
\bibinfo{author}{\bibfnamefont{R.}~\bibnamefont{Kubo}}, \bibinfo{journal}{Phys.
  Rev.} \textbf{\bibinfo{volume}{87}}, \bibinfo{pages}{568}
  (\bibinfo{year}{1952}).

\bibitem[{\citenamefont{Oguchi}(1960)}]{OguchPR60}
\bibinfo{author}{\bibfnamefont{T.}~\bibnamefont{Oguchi}},
  \bibinfo{journal}{Phys. Rev.} \textbf{\bibinfo{volume}{117}},
  \bibinfo{pages}{117} (\bibinfo{year}{1960}).

\bibitem[{\citenamefont{Takahashi}(1987)}]{TakahPRL87}
\bibinfo{author}{\bibfnamefont{M.}~\bibnamefont{Takahashi}},
  \bibinfo{journal}{Phys. Rev. Lett.} \textbf{\bibinfo{volume}{58}},
  \bibinfo{pages}{168} (\bibinfo{year}{1987}).

\bibitem[{\citenamefont{Takahashi}(1989)}]{TakahPRB89}
\bibinfo{author}{\bibfnamefont{M.}~\bibnamefont{Takahashi}},
  \bibinfo{journal}{Phys. Rev. B} \textbf{\bibinfo{volume}{40}},
  \bibinfo{pages}{2494} (\bibinfo{year}{1989}).

\bibitem[{\citenamefont{Hirsch and Tang}(1989)}]{HirscPRB89}
\bibinfo{author}{\bibfnamefont{J.~E.} \bibnamefont{Hirsch}} \bibnamefont{and}
  \bibinfo{author}{\bibfnamefont{S.}~\bibnamefont{Tang}},
  \bibinfo{journal}{Phys. Rev. B} \textbf{\bibinfo{volume}{40}},
  \bibinfo{pages}{4769} (\bibinfo{year}{1989}).

\bibitem[{\citenamefont{Tang et~al.}(1989)\citenamefont{Tang, Lazzouni, and
  Hirsch}}]{TangPRB89}
\bibinfo{author}{\bibfnamefont{S.}~\bibnamefont{Tang}},
  \bibinfo{author}{\bibfnamefont{M.~E.} \bibnamefont{Lazzouni}},
  \bibnamefont{and} \bibinfo{author}{\bibfnamefont{J.~E.}
  \bibnamefont{Hirsch}}, \bibinfo{journal}{Phys. Rev. B}
  \textbf{\bibinfo{volume}{40}}, \bibinfo{pages}{5000} (\bibinfo{year}{1989}).

\bibitem[{\citenamefont{Rezende}(1990)}]{RezenPRB90}
\bibinfo{author}{\bibfnamefont{S.~M.} \bibnamefont{Rezende}},
  \bibinfo{journal}{Phys. Rev. B} \textbf{\bibinfo{volume}{42}},
  \bibinfo{pages}{2589} (\bibinfo{year}{1990}).

\bibitem[{\citenamefont{Zhong and Sorella}(1993)}]{ZhongEL93}
\bibinfo{author}{\bibfnamefont{Q.~F.} \bibnamefont{Zhong}} \bibnamefont{and}
  \bibinfo{author}{\bibfnamefont{S.}~\bibnamefont{Sorella}},
  \bibinfo{journal}{Europhys. Lett.} \textbf{\bibinfo{volume}{21}},
  \bibinfo{pages}{629} (\bibinfo{year}{1993}).

\bibitem[{\citenamefont{Lavalle et~al.}(1998)\citenamefont{Lavalle, Sorella,
  and Parola}}]{LavalPRL98}
\bibinfo{author}{\bibfnamefont{C.}~\bibnamefont{Lavalle}},
  \bibinfo{author}{\bibfnamefont{S.}~\bibnamefont{Sorella}}, \bibnamefont{and}
  \bibinfo{author}{\bibfnamefont{A.}~\bibnamefont{Parola}},
  \bibinfo{journal}{Phys. Rev. Lett.} \textbf{\bibinfo{volume}{80}},
  \bibinfo{pages}{1746} (\bibinfo{year}{1998}).

\bibitem[{\citenamefont{Trumper et~al.}(2000)\citenamefont{Trumper, Capriotti,
  and Sorella}}]{TrumpPRB00}
\bibinfo{author}{\bibfnamefont{A.~E.} \bibnamefont{Trumper}},
  \bibinfo{author}{\bibfnamefont{L.}~\bibnamefont{Capriotti}},
  \bibnamefont{and} \bibinfo{author}{\bibfnamefont{S.}~\bibnamefont{Sorella}},
  \bibinfo{journal}{Phys. Rev. B} \textbf{\bibinfo{volume}{61}},
  \bibinfo{pages}{11529} (\bibinfo{year}{2000}).

\bibitem[{\citenamefont{Itoh et~al.}(1995)\citenamefont{Itoh, Endoh, Kakurai,
  and Tanaka}}]{ItohPRL95}
\bibinfo{author}{\bibfnamefont{S.}~\bibnamefont{Itoh}},
  \bibinfo{author}{\bibfnamefont{Y.}~\bibnamefont{Endoh}},
  \bibinfo{author}{\bibfnamefont{K.}~\bibnamefont{Kakurai}}, \bibnamefont{and}
  \bibinfo{author}{\bibfnamefont{H.}~\bibnamefont{Tanaka}},
  \bibinfo{journal}{Phys. Rev. Lett.} \textbf{\bibinfo{volume}{74}},
  \bibinfo{pages}{2375} (\bibinfo{year}{1995}).

\bibitem[{\citenamefont{Ivanov and Ivanov}(1992)}]{IvanoPRB92}
\bibinfo{author}{\bibfnamefont{N.~E.} \bibnamefont{Ivanov}} \bibnamefont{and}
  \bibinfo{author}{\bibfnamefont{P.~C.} \bibnamefont{Ivanov}},
  \bibinfo{journal}{Phys. Rev. B} \textbf{\bibinfo{volume}{46}},
  \bibinfo{pages}{8206} (\bibinfo{year}{1992}).

\bibitem[{\citenamefont{Wieser et~al.}(2008)\citenamefont{Wieser, Vedmedenko,
  and Wiesendanger}}]{WiesePRL08}
\bibinfo{author}{\bibfnamefont{R.}~\bibnamefont{Wieser}},
  \bibinfo{author}{\bibfnamefont{E.~Y.} \bibnamefont{Vedmedenko}},
  \bibnamefont{and}
  \bibinfo{author}{\bibfnamefont{R.}~\bibnamefont{Wiesendanger}},
  \bibinfo{journal}{Phys. Rev. Lett.} \textbf{\bibinfo{volume}{101}},
  \bibinfo{eid}{177202}  (\bibinfo{year}{2008}).

\bibitem[{\citenamefont{Ochsenbein et~al.}(2007)\citenamefont{Ochsenbein,
  Waldmann, Sieber, Carver, Bircher, G\"udel, Davies, Timco, Winpenny, Mutka, and Fernandez-Alonso}}]{OchseEL07}
\bibinfo{author}{\bibfnamefont{S.~T.} \bibnamefont{Ochsenbein}},
  \bibinfo{author}{\bibfnamefont{O.}~\bibnamefont{Waldmann}},
  \bibinfo{author}{\bibfnamefont{A.}~\bibnamefont{Sieber}},
  \bibinfo{author}{\bibfnamefont{G.}~\bibnamefont{Carver}},
  \bibinfo{author}{\bibfnamefont{R.}~\bibnamefont{Bircher}},
  \bibinfo{author}{\bibfnamefont{H.~U.} \bibnamefont{G\"udel}},
  \bibinfo{author}{\bibfnamefont{R.~S.~G.} \bibnamefont{Davies}},
  \bibinfo{author}{\bibfnamefont{G.~A.} \bibnamefont{Timco}},
  \bibinfo{author}{\bibfnamefont{R.~E.~P.} \bibnamefont{Winpenny}},
  \bibinfo{author}{\bibfnamefont{H.}~\bibnamefont{Mutka}}, \bibnamefont{and}
  \bibinfo{author}{\bibfnamefont{F.}~\bibnamefont{Fernandez-Alonso}}, 
  \bibinfo{journal}{Europhys. Lett.}
  \textbf{\bibinfo{volume}{79}}, \bibinfo{pages}{17003}
  (\bibinfo{year}{2007}).

\bibitem[{\citenamefont{Ochsenbein et~al.}(2008)\citenamefont{Ochsenbein, Tuna,
  Rancan, Davies, Muryn, Waldmann, Bircher, Sieber, Carver, Mutka, Fernandez-Alonso, Podlesnyak, Engelhardt, Timco, G\"udel, 
  and Winpenny}}]{OchseCEJ08}
\bibinfo{author}{\bibfnamefont{S.~T.} \bibnamefont{Ochsenbein}},
  \bibinfo{author}{\bibfnamefont{F.}~\bibnamefont{Tuna}},
  \bibinfo{author}{\bibfnamefont{M.}~\bibnamefont{Rancan}},
  \bibinfo{author}{\bibfnamefont{R.~S.~G.} \bibnamefont{Davies}},
  \bibinfo{author}{\bibfnamefont{C.~A.} \bibnamefont{Muryn}},
  \bibinfo{author}{\bibfnamefont{O.}~\bibnamefont{Waldmann}},
  \bibinfo{author}{\bibfnamefont{R.}~\bibnamefont{Bircher}},
  \bibinfo{author}{\bibfnamefont{A.}~\bibnamefont{Sieber}},
  \bibinfo{author}{\bibfnamefont{G.}~\bibnamefont{Carver}},
  \bibinfo{author}{\bibfnamefont{H.}~\bibnamefont{Mutka}},
  \bibinfo{author}{\bibfnamefont{F.}~\bibnamefont{Fernandez-Alonso}},
  \bibinfo{author}{\bibfnamefont{A.}~\bibnamefont{Podlesnyak}},
  \bibinfo{author}{\bibfnamefont{L.~P.}~\bibnamefont{Engelhardt}},
  \bibinfo{author}{\bibfnamefont{G.~A.}~\bibnamefont{Timco}},
  \bibinfo{author}{\bibfnamefont{H.~U.}~\bibnamefont{G\"udel}}, \bibnamefont{and}
  \bibinfo{author}{\bibfnamefont{R.~E.~P.} \bibnamefont{Winpenny}},
  \bibinfo{journal}{Chem. Eur. J.}
  \textbf{\bibinfo{volume}{14}}, \bibinfo{pages}{5144} (\bibinfo{year}{2008}).

\bibitem[{\citenamefont{Chaboussant et~al.}(2004)\citenamefont{Chaboussant,
  Sieber, Ochsenbein, G\"udel, Murrie, Honecker, Fukushima, and
  Normand}}]{ChaboPRB04}
\bibinfo{author}{\bibfnamefont{G.}~\bibnamefont{Chaboussant}},
  \bibinfo{author}{\bibfnamefont{A.}~\bibnamefont{Sieber}},
  \bibinfo{author}{\bibfnamefont{S.}~\bibnamefont{Ochsenbein}},
  \bibinfo{author}{\bibfnamefont{H.-U.} \bibnamefont{G\"udel}},
  \bibinfo{author}{\bibfnamefont{M.}~\bibnamefont{Murrie}},
  \bibinfo{author}{\bibfnamefont{A.}~\bibnamefont{Honecker}},
  \bibinfo{author}{\bibfnamefont{N.}~\bibnamefont{Fukushima}},
  \bibnamefont{and} \bibinfo{author}{\bibfnamefont{B.}~\bibnamefont{Normand}},
  \bibinfo{journal}{Phys. Rev. B} \textbf{\bibinfo{volume}{70}},
  \bibinfo{pages}{104422} (\bibinfo{year}{2004}).

\bibitem[{\citenamefont{Waldmann et~al.}(2003)\citenamefont{Waldmann, Guidi,
  Carretta, Mondelli, and Dearden}}]{WaldmPRL03}
\bibinfo{author}{\bibfnamefont{O.}~\bibnamefont{Waldmann}},
  \bibinfo{author}{\bibfnamefont{T.}~\bibnamefont{Guidi}},
  \bibinfo{author}{\bibfnamefont{S.}~\bibnamefont{Carretta}},
  \bibinfo{author}{\bibfnamefont{C.}~\bibnamefont{Mondelli}}, \bibnamefont{and}
  \bibinfo{author}{\bibfnamefont{A.~L.} \bibnamefont{Dearden}},
  \bibinfo{journal}{Phys. Rev. Lett.} \textbf{\bibinfo{volume}{91}},
  \bibinfo{eid}{237202} (\bibinfo{year}{2003}).

\bibitem[{\citenamefont{Yamamoto and Nakanishi}(2002)\citenamefont{Yamamoto, and Nakanishi}}]{YamamPRL02}
\bibinfo{author}{\bibfnamefont{S.}~\bibnamefont{Yamamoto}} \bibnamefont{and}
  \bibinfo{author}{\bibfnamefont{T.}~\bibnamefont{Nakanishi}},
  \bibinfo{journal}{Phys. Rev. Lett.}
  \textbf{\bibinfo{volume}{89}}, \bibinfo{pages}{157603} (\bibinfo{year}{2002}).


\bibitem[{\citenamefont{Hori and Yamamoto}(2003)\citenamefont{Hori, and Yamamoto}}]{HoriPRB03}
\bibinfo{author}{\bibfnamefont{H.}~\bibnamefont{Hori}} \bibnamefont{and}
  \bibinfo{author}{\bibfnamefont{S.}~\bibnamefont{Yamamoto}},
  \bibinfo{journal}{Phys. Rev. B}
  \textbf{\bibinfo{volume}{68}}, \bibinfo{pages}{054409} (\bibinfo{year}{2003}).



\bibitem[{\citenamefont{D.~Gatteschi}(2006)}]{Dante06}
\bibinfo{author}{\bibfnamefont{D.~Gatteschi} \bibnamefont{and} \bibfnamefont{R.~Sessoli}}, \emph{\bibinfo{title}{Molecular
  Nanomagnets}} (\bibinfo{publisher}{Oxford University Press},
  \bibinfo{year}{2006}).

\bibitem[{\citenamefont{Barbara et~al.}(1999)\citenamefont{Barbara, Thomas,
  Lionti, Chiorescu, and Sulpice}}]{BarbaJoMaMM99}
\bibinfo{author}{\bibfnamefont{B.}~\bibnamefont{Barbara}},
  \bibinfo{author}{\bibfnamefont{L.}~\bibnamefont{Thomas}},
  \bibinfo{author}{\bibfnamefont{F.}~\bibnamefont{Lionti}},
  \bibinfo{author}{\bibfnamefont{I.}~\bibnamefont{Chiorescu}},
  \bibnamefont{and} \bibinfo{author}{\bibfnamefont{A.}~\bibnamefont{Sulpice}},
  \bibinfo{journal}{J. Magn. Magn. Mater.}
  \textbf{\bibinfo{volume}{200}}, \bibinfo{pages}{167} (\bibinfo{year}{1999}).

\bibitem[{\citenamefont{Christou and Hendrickson}(2000)}]{GChMB00}
\bibinfo{author}{\bibfnamefont{G.~D.} \bibnamefont{Christou}},
  \bibnamefont{and} \bibinfo{author}{\bibfnamefont{D.~N.}
  \bibnamefont{Hendrickson}}, \bibinfo{journal}{MRS Bull.}
  \textbf{\bibinfo{volume}{25}}, \bibinfo{pages}{66} (\bibinfo{year}{2000}).

\bibitem[{\citenamefont{Gatteschi and Sessoli}(2003)}]{GatteACIE03}
\bibinfo{author}{\bibfnamefont{D.}~\bibnamefont{Gatteschi}} \bibnamefont{and}
  \bibinfo{author}{\bibfnamefont{R.}~\bibnamefont{Sessoli}},
  \bibinfo{journal}{Angew. Chem. Int. Ed.}
  \textbf{\bibinfo{volume}{42}}, \bibinfo{pages}{268} (\bibinfo{year}{2003}).


\bibitem[{\citenamefont{Wernsdorfer}(2007)}]{Werns07}
\bibinfo{author}{\bibfnamefont{W.}~\bibnamefont{Wernsdorfer}}, 
\bibinfo{journal}{Adv. Chem. Phys.} \textbf{\bibinfo{volume}{118}}, \bibinfo{pages}{99}
  (\bibinfo{year}{2007}).


\bibitem[{\citenamefont{Taft et~al.}(1994)\citenamefont{Taft, Delfs,
  Papaefthymiou, Foner, Gatteschi, and Lippard}}]{TaftJotACS94}
\bibinfo{author}{\bibfnamefont{K.~L.} \bibnamefont{Taft}},
  \bibinfo{author}{\bibfnamefont{C.~D.} \bibnamefont{Delfs}},
  \bibinfo{author}{\bibfnamefont{G.~C.} \bibnamefont{Papaefthymiou}},
  \bibinfo{author}{\bibfnamefont{S.}~\bibnamefont{Foner}},
  \bibinfo{author}{\bibfnamefont{D.}~\bibnamefont{Gatteschi}},
  \bibnamefont{and} \bibinfo{author}{\bibfnamefont{S.~J.}
  \bibnamefont{Lippard}}, \bibinfo{journal}{J. Am. Chem. Soc.} \textbf{\bibinfo{volume}{116}}, \bibinfo{pages}{823}
  (\bibinfo{year}{1994}).

\bibitem[{\citenamefont{Julien et~al.}(1999)\citenamefont{Julien, Jang,
  Lascialfari, Borsa, Horvatic, Caneschi, and
  Gatteschi}}]{JuliePRL99}
\bibinfo{author}{\bibfnamefont{M.-H.} \bibnamefont{Julien}},
  \bibinfo{author}{\bibfnamefont{Z.~H.} \bibnamefont{Jang}},
  \bibinfo{author}{\bibfnamefont{A.}~\bibnamefont{Lascialfari}},
  \bibinfo{author}{\bibfnamefont{F.}~\bibnamefont{Borsa}},
  \bibinfo{author}{\bibfnamefont{M.}~\bibnamefont{Horvatic}}, 
  \bibinfo{author}{\bibfnamefont{A.}~\bibnamefont{Caneschi}},
  \bibnamefont{and}
  \bibinfo{author}{\bibfnamefont{D.}~\bibnamefont{Gatteschi}},
  \bibinfo{journal}{Phys. Rev. Lett.} \textbf{\bibinfo{volume}{83}},
  \bibinfo{pages}{227} (\bibinfo{year}{1999}).

\bibitem[{\citenamefont{Cornia et~al.}(1999)\citenamefont{Cornia, Jansen, and
  Affronte}}]{CorniPRB99}
\bibinfo{author}{\bibfnamefont{A.}~\bibnamefont{Cornia}},
  \bibinfo{author}{\bibfnamefont{A.~G.~M.} \bibnamefont{Jansen}},
  \bibnamefont{and} \bibinfo{author}{\bibfnamefont{M.}~\bibnamefont{Affronte}},
  \bibinfo{journal}{Phys. Rev. B} \textbf{\bibinfo{volume}{60}},
  \bibinfo{pages}{12177} (\bibinfo{year}{1999}).

\bibitem[{\citenamefont{van Slageren et~al.}(2004)\citenamefont{van Slageren, Sessoli, Gatteschi, Smith, Helliwell, Winpenny, Cornia, Barra, Jansen, Rentschler, and Timco}}]{SlageC-AEJ02}
  \bibinfo{author}{\bibfnamefont{J.}~\bibnamefont{van Slageren}},
  \bibinfo{author}{\bibfnamefont{R.}~\bibnamefont{Sessoli}},
  \bibinfo{author}{\bibfnamefont{D.}~\bibnamefont{Gatteschi}},
  \bibinfo{author}{\bibfnamefont{A.~A.}~\bibnamefont{Smith}},
  \bibinfo{author}{\bibfnamefont{M.}~\bibnamefont{Helliwell}},
  \bibinfo{author}{\bibfnamefont{R.~E.~P.}~\bibnamefont{Winpenny}},
  \bibinfo{author}{\bibfnamefont{A.}~\bibnamefont{Cornia}},
  \bibinfo{author}{\bibfnamefont{A.~L.}~\bibnamefont{Barra}},
  \bibinfo{author}{\bibfnamefont{A.~G.~M.}~\bibnamefont{Jansen}},
  \bibinfo{author}{\bibfnamefont{E.}~\bibnamefont{Rentschler}},
 \bibnamefont{and}  \bibinfo{author}{\bibfnamefont{G.~A.}~\bibnamefont{Timco}}, 
  \bibinfo{journal}{Chem. Eur. J.} \textbf{\bibinfo{volume}{8}},
    \bibinfo{pages}{277} (\bibinfo{year}{2002}).

\bibitem[{\citenamefont{Carretta et~al.}(2003)\citenamefont{Carretta, van
  Slageren, Guidi, Liviotti, Mondelli, Rovai, Cornia, Dearden, Carsughi,
  Affronte, Frost, Winpenny, Gatteschi, Amoretti, and Caciuffo}}]{CarrePRB03}
\bibinfo{author}{\bibfnamefont{S.}~\bibnamefont{Carretta}},
  \bibinfo{author}{\bibfnamefont{J.}~\bibnamefont{van Slageren}},
  \bibinfo{author}{\bibfnamefont{T.}~\bibnamefont{Guidi}},
  \bibinfo{author}{\bibfnamefont{E.}~\bibnamefont{Liviotti}},
  \bibinfo{author}{\bibfnamefont{C.}~\bibnamefont{Mondelli}},
  \bibinfo{author}{\bibfnamefont{D.}~\bibnamefont{Rovai}},
  \bibinfo{author}{\bibfnamefont{A.}~\bibnamefont{Cornia}},
  \bibinfo{author}{\bibfnamefont{A.~L.} \bibnamefont{Dearden}},
  \bibinfo{author}{\bibfnamefont{F.}~\bibnamefont{Carsughi}},
  \bibinfo{author}{\bibfnamefont{M.}~\bibnamefont{Affronte}},
  \bibinfo{author}{\bibfnamefont{C.~D.}~\bibnamefont{Frost}},
  \bibinfo{author}{\bibfnamefont{R.~E.~P.}~\bibnamefont{Winpenny}},
  \bibinfo{author}{\bibfnamefont{D.}~\bibnamefont{Gatteschi}},
  \bibinfo{author}{\bibfnamefont{G.}~\bibnamefont{Amoretti}},  \bibnamefont{and}
  \bibinfo{author}{\bibfnamefont{R.}~\bibnamefont{Caciuffo}},
  \bibinfo{journal}{Phys. Rev. B}
  \textbf{\bibinfo{volume}{67}}, \bibinfo{pages}{094405}
  (\bibinfo{year}{2003}).

\bibitem[{\citenamefont{Pilawa et~al.}(2005)\citenamefont{Pilawa, Boffinger,
  Keilhauer, Leppin, Odenwald, Wendl, Berthier, and
  Horvatic}}]{PilawPRB05}
\bibinfo{author}{\bibfnamefont{B.}~\bibnamefont{Pilawa}},
  \bibinfo{author}{\bibfnamefont{R.}~\bibnamefont{Boffinger}},
  \bibinfo{author}{\bibfnamefont{I.}~\bibnamefont{Keilhauer}},
  \bibinfo{author}{\bibfnamefont{R.}~\bibnamefont{Leppin}},
  \bibinfo{author}{\bibfnamefont{I.}~\bibnamefont{Odenwald}},
  \bibinfo{author}{\bibfnamefont{W.}~\bibnamefont{Wendl}},
  \bibinfo{author}{\bibfnamefont{C.}~\bibnamefont{Berthier}}, \bibnamefont{and}
  \bibinfo{author}{\bibfnamefont{M.}~\bibnamefont{Horvatic}}, 
  \bibinfo{journal}{Phys. Rev. B} \textbf{\bibinfo{volume}{71}},
  \bibinfo{pages}{184419} (\bibinfo{year}{2005}).

\bibitem[{\citenamefont{Anderson}(1984)}]{Ander84}
\bibinfo{author}{\bibfnamefont{P.~W.} \bibnamefont{Anderson}},
  \emph{\bibinfo{title}{Basic Notions of Condensed Matter Physics}}
  (\bibinfo{publisher}{Benjamin/Cummings, Menlo Park},
  \bibinfo{year}{1984}).

\bibitem[{\citenamefont{Bernu et~al.}(1992)\citenamefont{Bernu, Lhuillier, and
  Pierre}}]{BernuPRL92}
\bibinfo{author}{\bibfnamefont{B.}~\bibnamefont{Bernu}},
  \bibinfo{author}{\bibfnamefont{C.}~\bibnamefont{Lhuillier}},
  \bibnamefont{and} \bibinfo{author}{\bibfnamefont{L.}~\bibnamefont{Pierre}},
  \bibinfo{journal}{Phys. Rev. Lett.} \textbf{\bibinfo{volume}{69}},
  \bibinfo{pages}{2590} (\bibinfo{year}{1992}).

\bibitem[{\citenamefont{Bernu et~al.}(1994)\citenamefont{Bernu, Lecheminant,
  Lhuillier, and Pierre}}]{BernuPRB94}
\bibinfo{author}{\bibfnamefont{B.}~\bibnamefont{Bernu}},
  \bibinfo{author}{\bibfnamefont{P.}~\bibnamefont{Lecheminant}},
  \bibinfo{author}{\bibfnamefont{C.}~\bibnamefont{Lhuillier}},
  \bibnamefont{and} \bibinfo{author}{\bibfnamefont{L.}~\bibnamefont{Pierre}},
  \bibinfo{journal}{Phys. Rev. B} \textbf{\bibinfo{volume}{50}},
  \bibinfo{pages}{10048} (\bibinfo{year}{1994}).

\bibitem[{\citenamefont{Chiolero and Loss}(1998)}]{ChiolPRL98}
\bibinfo{author}{\bibfnamefont{A.}~\bibnamefont{Chiolero}} \bibnamefont{and}
  \bibinfo{author}{\bibfnamefont{D.}~\bibnamefont{Loss}},
  \bibinfo{journal}{Phys. Rev. Lett.} \textbf{\bibinfo{volume}{80}},
  \bibinfo{pages}{169} (\bibinfo{year}{1998}).

\bibitem[{\citenamefont{Schnack and Luban}(2000)}]{SchnaPRB00}
\bibinfo{author}{\bibfnamefont{J.}~\bibnamefont{Schnack}} \bibnamefont{and}
  \bibinfo{author}{\bibfnamefont{M.}~\bibnamefont{Luban}},
  \bibinfo{journal}{Phys. Rev. B} \textbf{\bibinfo{volume}{63}},
  \bibinfo{pages}{014418} (\bibinfo{year}{2000}).

\bibitem[{\citenamefont{Waldmann}(2001)}]{WaldmPRB01}
\bibinfo{author}{\bibfnamefont{O.}~\bibnamefont{Waldmann}},
  \bibinfo{journal}{Phys. Rev. B} \textbf{\bibinfo{volume}{65}},
  \bibinfo{pages}{024424} (\bibinfo{year}{2001}).

\bibitem[{\citenamefont{Lhuillier}(2005)}]{Lhuil05}
\bibinfo{author}{\bibfnamefont{C.}~\bibnamefont{Lhuillier}},
  \bibinfo{journal}{arXiv:cond-mat/0502464}
  (\bibinfo{year}{2005}).

\bibitem[{\citenamefont{Guidi et~al.}(2004)\citenamefont{Guidi, Carretta,
  Santini, Liviotti, Magnani, Mondelli, Waldmann, Thompson, Zhao, Frost, Amoretti, and Caciuffo}}]{GuidiPRB04}
\bibinfo{author}{\bibfnamefont{T.}~\bibnamefont{Guidi}},
  \bibinfo{author}{\bibfnamefont{S.}~\bibnamefont{Carretta}},
  \bibinfo{author}{\bibfnamefont{P.}~\bibnamefont{Santini}},
  \bibinfo{author}{\bibfnamefont{E.}~\bibnamefont{Liviotti}},
  \bibinfo{author}{\bibfnamefont{N.}~\bibnamefont{Magnani}},
  \bibinfo{author}{\bibfnamefont{C.}~\bibnamefont{Mondelli}},
  \bibinfo{author}{\bibfnamefont{O.}~\bibnamefont{Waldmann}},
  \bibinfo{author}{\bibfnamefont{L.~K.} \bibnamefont{Thompson}},
  \bibinfo{author}{\bibfnamefont{L.}~\bibnamefont{Zhao}},
  \bibinfo{author}{\bibfnamefont{C.~D.} \bibnamefont{Frost}},
  \bibinfo{author}{\bibfnamefont{G.}~\bibnamefont{Amoretti}}, \bibnamefont{and}
  \bibinfo{author}{\bibfnamefont{R.}~\bibnamefont{Caciuffo}},
  \bibinfo{journal}{Phys. Rev. B}
  \textbf{\bibinfo{volume}{69}}, \bibinfo{pages}{104432}
  (\bibinfo{year}{2004}).

\bibitem[{\citenamefont{Caciuffo et~al.}(2005)\citenamefont{Caciuffo, Guidi,
  Amoretti, Carretta, Liviotti, Santini, Mondelli, Timco, Muryn, and
  Winpenny}}]{CaciuPRB05}
\bibinfo{author}{\bibfnamefont{R.}~\bibnamefont{Caciuffo}},
  \bibinfo{author}{\bibfnamefont{T.}~\bibnamefont{Guidi}},
  \bibinfo{author}{\bibfnamefont{G.}~\bibnamefont{Amoretti}},
  \bibinfo{author}{\bibfnamefont{S.}~\bibnamefont{Carretta}},
  \bibinfo{author}{\bibfnamefont{E.}~\bibnamefont{Liviotti}},
  \bibinfo{author}{\bibfnamefont{P.}~\bibnamefont{Santini}},
  \bibinfo{author}{\bibfnamefont{C.}~\bibnamefont{Mondelli}},
  \bibinfo{author}{\bibfnamefont{G.}~\bibnamefont{Timco}},
  \bibinfo{author}{\bibfnamefont{C.~A.} \bibnamefont{Muryn}}, \bibnamefont{and}
  \bibinfo{author}{\bibfnamefont{R.~E.~P.} \bibnamefont{Winpenny}},
  \bibinfo{journal}{Phys. Rev. B} \textbf{\bibinfo{volume}{71}},
  \bibinfo{pages}{174407} (\bibinfo{year}{2005}).

\bibitem[{\citenamefont{Waldmann}(2005)}]{WaldmCCR05}
\bibinfo{author}{\bibfnamefont{O.}~\bibnamefont{Waldmann}},
  \bibinfo{journal}{Coord. Chem. Rev.}
  \textbf{\bibinfo{volume}{249}}, \bibinfo{pages}{2550} (\bibinfo{year}{2005}).

\bibitem[{\citenamefont{Santini et~al.}(2005)\citenamefont{Santini, Carretta,
  Amoretti, Guidi, Caciuffo, Caneschi, Rovai, Qiu, and Copley}}]{SantiPRB05}
\bibinfo{author}{\bibfnamefont{P.}~\bibnamefont{Santini}},
  \bibinfo{author}{\bibfnamefont{S.}~\bibnamefont{Carretta}},
  \bibinfo{author}{\bibfnamefont{G.}~\bibnamefont{Amoretti}},
  \bibinfo{author}{\bibfnamefont{T.}~\bibnamefont{Guidi}},
  \bibinfo{author}{\bibfnamefont{R.}~\bibnamefont{Caciuffo}},
  \bibinfo{author}{\bibfnamefont{A.}~\bibnamefont{Caneschi}},
  \bibinfo{author}{\bibfnamefont{D.}~\bibnamefont{Rovai}},
  \bibinfo{author}{\bibfnamefont{Y.}~\bibnamefont{Qiu}}, \bibnamefont{and}
  \bibinfo{author}{\bibfnamefont{J.~R.~D.} \bibnamefont{Copley}},
  \bibinfo{journal}{Phys. Rev. B} \textbf{\bibinfo{volume}{71}},
  \bibinfo{pages}{184405} (\bibinfo{year}{2005}).

\bibitem[{\citenamefont{Waldmann
  et~al.}(2006{\natexlab{a}})\citenamefont{Waldmann, G\"udel, Kelly, and
  Thompson}}]{WaldmIC06}
\bibinfo{author}{\bibfnamefont{O.}~\bibnamefont{Waldmann}},
  \bibinfo{author}{\bibfnamefont{H.~U.} \bibnamefont{G\"udel}},
  \bibinfo{author}{\bibfnamefont{T.~L.} \bibnamefont{Kelly}}, \bibnamefont{and}
  \bibinfo{author}{\bibfnamefont{L.~K.} \bibnamefont{Thompson}},
  \bibinfo{journal}{Inorg. Chem.} \textbf{\bibinfo{volume}{45}},
  \bibinfo{pages}{3295} (\bibinfo{year}{2006}{\natexlab{a}}).

\bibitem[{\citenamefont{Guidi et~al.}(2007)\citenamefont{Guidi, Copley, Qiu,
  Carretta, Santini, Amoretti, Timco, Winpenny, Dennis, and
  Caciuffo}}]{GuidiPRB07}
\bibinfo{author}{\bibfnamefont{T.}~\bibnamefont{Guidi}},
  \bibinfo{author}{\bibfnamefont{J.~R.~D.} \bibnamefont{Copley}},
  \bibinfo{author}{\bibfnamefont{Y.}~\bibnamefont{Qiu}},
  \bibinfo{author}{\bibfnamefont{S.}~\bibnamefont{Carretta}},
  \bibinfo{author}{\bibfnamefont{P.}~\bibnamefont{Santini}},
  \bibinfo{author}{\bibfnamefont{G.}~\bibnamefont{Amoretti}},
  \bibinfo{author}{\bibfnamefont{G.}~\bibnamefont{Timco}},
  \bibinfo{author}{\bibfnamefont{R.~E.~P.} \bibnamefont{Winpenny}},
  \bibinfo{author}{\bibfnamefont{C.~L.} \bibnamefont{Dennis}},
  \bibnamefont{and} \bibinfo{author}{\bibfnamefont{R.}~\bibnamefont{Caciuffo}},
  \bibinfo{journal}{Phys. Rev. B} \textbf{\bibinfo{volume}{75}},
  \bibinfo{pages}{014408} (\bibinfo{year}{2007}).
  
\bibitem[{\citenamefont{Schnack et~al.}(2001)\citenamefont{Schnack, Luban, and
  Modler}}]{SchnaEL01}
\bibinfo{author}{\bibfnamefont{J.}~\bibnamefont{Schnack}},
  \bibinfo{author}{\bibfnamefont{M.}~\bibnamefont{Luban}}, \bibnamefont{and}
  \bibinfo{author}{\bibfnamefont{R.}~\bibnamefont{Modler}},
  \bibinfo{journal}{Europhys. Lett.} \textbf{\bibinfo{volume}{56}},
  \bibinfo{pages}{863} (\bibinfo{year}{2001}).

\bibitem[{\citenamefont{Garlea et~al.}(2006)\citenamefont{Garlea, Nagler,
  Zarestky, Stassis, Vaknin, K\"{o}gerler, McMorrow, Niedermayer, Tennant, Lake, Qiu, Exler, Schnack, and Luban}}]{GarlePRB06}
\bibinfo{author}{\bibfnamefont{V.~O.} \bibnamefont{Garlea}},
  \bibinfo{author}{\bibfnamefont{S.~E.} \bibnamefont{Nagler}},
  \bibinfo{author}{\bibfnamefont{J.~L.} \bibnamefont{Zarestky}},
  \bibinfo{author}{\bibfnamefont{C.}~\bibnamefont{Stassis}},
  \bibinfo{author}{\bibfnamefont{D.}~\bibnamefont{Vaknin}},
  \bibinfo{author}{\bibfnamefont{P.}~\bibnamefont{K\"{o}gerler}},
  \bibinfo{author}{\bibfnamefont{D.~F.} \bibnamefont{McMorrow}},
  \bibinfo{author}{\bibfnamefont{C.}~\bibnamefont{Niedermayer}},
  \bibinfo{author}{\bibfnamefont{D.~A.} \bibnamefont{Tennant}},
  \bibinfo{author}{\bibfnamefont{B.}~\bibnamefont{Lake}}, 
  \bibinfo{author}{\bibfnamefont{Y.}~\bibnamefont{Qiu}}, 
  \bibinfo{author}{\bibfnamefont{M.}~\bibnamefont{Exler}}, 
  \bibinfo{author}{\bibfnamefont{J.}~\bibnamefont{Schnack}},  \bibnamefont{and}
  \bibinfo{author}{\bibfnamefont{M.}~\bibnamefont{Luban}},
  \bibinfo{journal}{Phys. Rev. B} \textbf{\bibinfo{volume}{73}},
  \bibinfo{eid}{024414} (\bibinfo{year}{2006}).

\bibitem[{\citenamefont{Waldmann}(2007)}]{WaldmPRB07b}
\bibinfo{author}{\bibfnamefont{O.}~\bibnamefont{Waldmann}},
  \bibinfo{journal}{Phys. Rev. B} \textbf{\bibinfo{volume}{75}},
  \bibinfo{eid}{012415} (\bibinfo{year}{2007}).

\bibitem[{\citenamefont{M\"uller et~al.}(1981)\citenamefont{M\"uller, Thomas,
  Beck, and Bonner}}]{MuellePRB81}
\bibinfo{author}{\bibfnamefont{G.}~\bibnamefont{M\"uller}},
  \bibinfo{author}{\bibfnamefont{H.}~\bibnamefont{Thomas}},
  \bibinfo{author}{\bibfnamefont{H.}~\bibnamefont{Beck}}, \bibnamefont{and}
  \bibinfo{author}{\bibfnamefont{J.~C.} \bibnamefont{Bonner}},
  \bibinfo{journal}{Phys. Rev. B} \textbf{\bibinfo{volume}{24}},
  \bibinfo{pages}{1429} (\bibinfo{year}{1981}).

\bibitem[{\citenamefont{Haldane}(1983)}]{HaldaPRL83}
\bibinfo{author}{\bibfnamefont{F.~D.~M.} \bibnamefont{Haldane}},
  \bibinfo{journal}{Phys. Rev. Lett.} \textbf{\bibinfo{volume}{50}},
  \bibinfo{pages}{1153} (\bibinfo{year}{1983}).

\bibitem[{\citenamefont{Schnack}(2000)}]{SchnaPRB00a}
\bibinfo{author}{\bibfnamefont{J.} \bibnamefont{Schnack}},
  \bibinfo{journal}{Phys. Rev. B} \textbf{\bibinfo{volume}{62}},
  \bibinfo{pages}{14855} (\bibinfo{year}{2000}).

\bibitem[{\citenamefont{Engelhardt and Luban}(2006)\citenamefont{Engelhardt, and Luban}}]{EngelPRB06}
\bibinfo{author}{\bibfnamefont{L.}~\bibnamefont{Engelhardt}} \bibnamefont{and}
  \bibinfo{author}{\bibfnamefont{M.}~\bibnamefont{Luban}},
  \bibinfo{journal}{Phys. Rev. B} \textbf{\bibinfo{volume}{73}},
  \bibinfo{pages}{054430} (\bibinfo{year}{2006}).

\bibitem[{\citenamefont{Saalfrank et~al.}(1997)\citenamefont{Saalfrank, Bernt,
  Uller, and Hampel}}]{SaalfACIEiE97}
\bibinfo{author}{\bibfnamefont{R.~W.} \bibnamefont{Saalfrank}},
  \bibinfo{author}{\bibfnamefont{I.}~\bibnamefont{Bernt}},
  \bibinfo{author}{\bibfnamefont{E.}~\bibnamefont{Uller}}, \bibnamefont{and}
  \bibinfo{author}{\bibfnamefont{F.}~\bibnamefont{Hampel}},
  \bibinfo{journal}{Angew. Chem., Int. Ed. Engl.}
  \textbf{\bibinfo{volume}{36}}, \bibinfo{pages}{2482} (\bibinfo{year}{1997}).

\bibitem[{\citenamefont{Waldmann et~al.}(2001)\citenamefont{Waldmann, Koch,
  Schromm, Sch\"ulein, M\"uller, Bernt, Saalfrank, Hampel, and
  Balthes}}]{WaldmIC01}
\bibinfo{author}{\bibfnamefont{O.}~\bibnamefont{Waldmann}},
  \bibinfo{author}{\bibfnamefont{R.}~\bibnamefont{Koch}},
  \bibinfo{author}{\bibfnamefont{S.}~\bibnamefont{Schromm}},
  \bibinfo{author}{\bibfnamefont{J.}~\bibnamefont{Sch\"ulein}},
  \bibinfo{author}{\bibfnamefont{P.}~\bibnamefont{M\"uller}},
  \bibinfo{author}{\bibfnamefont{I.}~\bibnamefont{Bernt}},
  \bibinfo{author}{\bibfnamefont{R.}~\bibnamefont{Saalfrank}},
  \bibinfo{author}{\bibfnamefont{F.}~\bibnamefont{Hampel}}, \bibnamefont{and}
  \bibinfo{author}{\bibfnamefont{E.}~\bibnamefont{Balthes}},
  \bibinfo{journal}{Inorg. Chem.} \textbf{\bibinfo{volume}{40}},
  \bibinfo{pages}{2986} (\bibinfo{year}{2001}).

\bibitem[{\citenamefont{Waldmann
  et~al.}(2006{\natexlab{b}})\citenamefont{Waldmann, Dobe, G\"udel, and
  Mutka}}]{WaldmPRB06}
\bibinfo{author}{\bibfnamefont{O.}~\bibnamefont{Waldmann}},
  \bibinfo{author}{\bibfnamefont{C.}~\bibnamefont{Dobe}},
  \bibinfo{author}{\bibfnamefont{H.~U.} \bibnamefont{G\"udel}}, \bibnamefont{and}
  \bibinfo{author}{\bibfnamefont{H.}~\bibnamefont{Mutka}},
  \bibinfo{journal}{Phys. Rev. B} \textbf{\bibinfo{volume}{74}},
  \bibinfo{eid}{054429} 
  (\bibinfo{year}{2006}{\natexlab{b}}).

\bibitem[{\citenamefont{Waldmann et~al.}(2005)\citenamefont{Waldmann, Dobe,
  Mutka, Furrer, and G\"udel}}]{WaldmPRL05}
\bibinfo{author}{\bibfnamefont{O.}~\bibnamefont{Waldmann}},
  \bibinfo{author}{\bibfnamefont{C.}~\bibnamefont{Dobe}},
  \bibinfo{author}{\bibfnamefont{H.}~\bibnamefont{Mutka}},
  \bibinfo{author}{\bibfnamefont{A.}~\bibnamefont{Furrer}}, \bibnamefont{and}
  \bibinfo{author}{\bibfnamefont{H.~U.} \bibnamefont{G\"udel}},
  \bibinfo{journal}{Phys. Rev. Lett.} \textbf{\bibinfo{volume}{95}},
  \bibinfo{eid}{057202} (\bibinfo{year}{2005}).

\bibitem[{\citenamefont{Waldmann}(2003)}]{WaldmPRB03}
\bibinfo{author}{\bibfnamefont{O.}~\bibnamefont{Waldmann}},
  \bibinfo{journal}{Phys. Rev. B} \textbf{\bibinfo{volume}{68}},
  \bibinfo{eid}{174406}  (\bibinfo{year}{2003}).

\bibitem[{\citenamefont{Shirane et~al.}(2002)\citenamefont{Shirane, Shapiro,
  and Tranquada}}]{Shira02}
\bibinfo{author}{\bibfnamefont{G.}~\bibnamefont{Shirane}},
  \bibinfo{author}{\bibfnamefont{S.~M.} \bibnamefont{Shapiro}},
  \bibnamefont{and} \bibinfo{author}{\bibfnamefont{J.~M.}
  \bibnamefont{Tranquada}}, \emph{\bibinfo{title}{Neutron Scattering with a
  Triple-Axis Spectrometer}} (\bibinfo{publisher}{Cambridge University Press, Cambridge, England},
  \bibinfo{year}{2002}).

\bibitem[{\citenamefont{Waldmann and G\"udel}(2005)}]{WaldmPRB05a}
\bibinfo{author}{\bibfnamefont{O.}~\bibnamefont{Waldmann}} \bibnamefont{and}
  \bibinfo{author}{\bibfnamefont{H.~U.} \bibnamefont{G\"udel}},
  \bibinfo{journal}{Phys. Rev. B} \textbf{\bibinfo{volume}{72}},
  \bibinfo{eid}{094422} (\bibinfo{year}{2005}).

\bibitem[{\citenamefont{Waldmann
  et~al.}(2006{\natexlab{c}})\citenamefont{Waldmann, Dobe, Ochsenbein, G\"udel,
  and Sheikin}}]{WaldmPRL06}
\bibinfo{author}{\bibfnamefont{O.}~\bibnamefont{Waldmann}},
  \bibinfo{author}{\bibfnamefont{C.}~\bibnamefont{Dobe}},
  \bibinfo{author}{\bibfnamefont{S.~T.} \bibnamefont{Ochsenbein}},
  \bibinfo{author}{\bibfnamefont{H.~U.} \bibnamefont{G\"udel}}, \bibnamefont{and}
  \bibinfo{author}{\bibfnamefont{I.}~\bibnamefont{Sheikin}},
  \bibinfo{journal}{Phys. Rev. Lett.} \textbf{\bibinfo{volume}{96}},
  \bibinfo{eid}{027206}   (\bibinfo{year}{2006}{\natexlab{c}}).

\bibitem[{\citenamefont{Soncini and Chibotaru}(2007)}]{SonciPRL07}
\bibinfo{author}{\bibfnamefont{A.}~\bibnamefont{Soncini}} \bibnamefont{and}
  \bibinfo{author}{\bibfnamefont{L.~F.} \bibnamefont{Chibotaru}},
  \bibinfo{journal}{Phys. Rev. Lett.} \textbf{\bibinfo{volume}{99}},
  \bibinfo{eid}{077204} (\bibinfo{year}{2007}).

\bibitem[{\citenamefont{Lante et~al.}(2009)\citenamefont{Lante, Rousochatzakis,
  Penc, Waldmann, and Mila}}]{LantePRB09}
\bibinfo{author}{\bibfnamefont{V.}~\bibnamefont{Lante}},
  \bibinfo{author}{\bibfnamefont{I.}~\bibnamefont{Rousochatzakis}},
  \bibinfo{author}{\bibfnamefont{K.}~\bibnamefont{Penc}},
  \bibinfo{author}{\bibfnamefont{O.}~\bibnamefont{Waldmann}}, \bibnamefont{and}
  \bibinfo{author}{\bibfnamefont{F.}~\bibnamefont{Mila}},
  \bibinfo{journal}{Phys. Rev. B} \textbf{\bibinfo{volume}{79}},
  \bibinfo{eid}{180412(R)}  (\bibinfo{year}{2009}).

\bibitem[{\citenamefont{Cinti et~al.}(2002)\citenamefont{Cinti, Affronte, and
  Jansen}}]{CintiTEPJB02}
\bibinfo{author}{\bibfnamefont{F.}~\bibnamefont{Cinti}},
  \bibinfo{author}{\bibfnamefont{M.}~\bibnamefont{Affronte}}, \bibnamefont{and}
  \bibinfo{author}{\bibfnamefont{A.}~\bibnamefont{Jansen}},
  \bibinfo{journal}{Eur. Phys. J. B}
  \textbf{\bibinfo{volume}{30}}, \bibinfo{pages}{461} (\bibinfo{year}{2002}).

\bibitem[{\citenamefont{C\'epas and Ziman}(2005)}]{C'epPoTPS05}
\bibinfo{author}{\bibfnamefont{O.}~\bibnamefont{C\'epas}} \bibnamefont{and}
  \bibinfo{author}{\bibfnamefont{T.}~\bibnamefont{Ziman}},
  \bibinfo{journal}{Prog. Theor. Phys. Suppl.}
  \textbf{\bibinfo{volume}{159}}, \bibinfo{pages}{280} (\bibinfo{year}{2005}).

\bibitem[{\citenamefont{Auerbach and Arovas}(1988)}]{AuerbPRL88}
\bibinfo{author}{\bibfnamefont{A.}~\bibnamefont{Auerbach}} \bibnamefont{and}
  \bibinfo{author}{\bibfnamefont{D.~P.} \bibnamefont{Arovas}},
  \bibinfo{journal}{Phys. Rev. Lett.} \textbf{\bibinfo{volume}{61}},
  \bibinfo{pages}{617} (\bibinfo{year}{1988}).

\bibitem[{\citenamefont{Sarker et~al.}(1989)\citenamefont{Sarker, Jayaprakash,
  Krishnamurthy, and Ma}}]{SarkePRB89}
\bibinfo{author}{\bibfnamefont{S.}~\bibnamefont{Sarker}},
  \bibinfo{author}{\bibfnamefont{C.}~\bibnamefont{Jayaprakash}},
  \bibinfo{author}{\bibfnamefont{H.~R.} \bibnamefont{Krishnamurthy}},
  \bibnamefont{and} \bibinfo{author}{\bibfnamefont{M.}~\bibnamefont{Ma}},
  \bibinfo{journal}{Phys. Rev. B} \textbf{\bibinfo{volume}{40}},
  \bibinfo{pages}{5028} (\bibinfo{year}{1989}).

\bibitem[{\citenamefont{Yamamoto and Hori}(2003)}]{YamamJotPSoJ03}
\bibinfo{author}{\bibfnamefont{S.}~\bibnamefont{Yamamoto}} \bibnamefont{and}
  \bibinfo{author}{\bibfnamefont{H.}~\bibnamefont{Hori}},
  \bibinfo{journal}{J. Phys. Soc. Jpn.}
  \textbf{\bibinfo{volume}{72}}, \bibinfo{pages}{769} (\bibinfo{year}{2003}).

\bibitem[{\citenamefont{Arovas and Auerbach}(1988)}]{ArovaPRB88}
\bibinfo{author}{\bibfnamefont{D.~P.} \bibnamefont{Arovas}} \bibnamefont{and}
  \bibinfo{author}{\bibfnamefont{A.}~\bibnamefont{Auerbach}},
  \bibinfo{journal}{Phys. Rev. B} \textbf{\bibinfo{volume}{38}},
  \bibinfo{pages}{316} (\bibinfo{year}{1988}).

\end{thebibliography}
\end{document}